\documentclass[sigconf,10pt]{acmart}

\setcopyright{rightsretained}

\acmDOI{10.1145/1122445.1122456}
\acmISBN{978-1-4503-XXXX-X/18/06}
\acmConference[SoCC '23]{ACM Symposium on Cloud Computing}{October
30--November 1, 2023}{Santa Cruz, CA, USA}
\acmBooktitle{ACM Symposium on Cloud Computing (SoCC '23), October
30--November 1, 2023, Santa Cruz, CA, USA}
\copyrightyear{2023}
\acmYear{2023}

\usepackage[utf8]{inputenc}
\usepackage{graphicx}
\usepackage{booktabs} 
\usepackage{pifont}
\usepackage{bbold}

\usepackage{hyperref}
\usepackage{tikz}
\usepackage{amsmath}
\usepackage[]{subfig}

\usepackage[T1]{fontenc}
\usepackage{filecontents}
\usepackage{multirow}
\usepackage{graphicx}
\usepackage{booktabs} 
\usepackage{amsmath}
\usepackage{bm}
\usepackage{enumerate}
\usepackage{listings}
\usepackage{mathtools}
\usepackage{algorithm}
\usepackage{algorithmic}
\usepackage{float}
\usepackage{colortbl}
\usepackage{hyperref}
\usepackage{cleveref}
\usepackage{xspace}
\usepackage{tikz}
\usepackage[most]{tcolorbox}
\usepackage{enumitem}  
\usepackage{bigstrut} 
\usepackage{adjustbox}
\usepackage{array}
\usepackage{booktabs}
\usepackage{multirow}

\usepackage{lipsum}
\usepackage{titlesec}
\usepackage{tabularx}
\newcolumntype{R}[2]{%
    >{\adjustbox{angle=#1,lap=\width-(#2)}\bgroup}%
    l%
    <{\egroup}%
}

\titlespacing\section{0pt}{5pt plus 1pt minus 1pt}{3pt plus 1pt minus 1pt}
\titlespacing\subsection{0pt}{1pt plus 1pt minus 1pt}{1pt plus 1pt minus 1pt}
\titlespacing\subsubsection{0pt}{1pt plus 1pt minus 1pt}{1pt plus 1pt minus 1pt}

\usepackage{tabularx}
\newcommand{\ourapproach}{\textsc{Cameo}\xspace}
\newcommand{\unicorn}{\textsc{Unicorn}\xspace}

\newcommand{\cello}{\textsc{cello}\xspace}
\newcommand{\restune}{\textsc{ResTune-w/o-ML}\xspace}
\newcommand{\restuner}{\textsc{ResTune}\xspace}
\newcommand{\modp}{\textsc{Mlperf object detection}\xspace}
\newcommand{\xavier}{\textsc{Xavier}\xspace}
\newcommand{\txtwo}{\textsc{TX2}\xspace}
\newcommand{\txone}{\textsc{TX1}\xspace}
\newcommand{\chameleon}{\textsc{Chameleon cloud instance}\xspace}
\newcommand{\deepstream}{\textsc{DeepStream}\xspace}
\newcommand{\smac}{\textsc{Smac}\xspace}

\newcommand\iftyA[1]{{\color{red}{citation\_needed\_maybe}}}
 
\usepackage{tikz}
\usepackage{amsmath}
\usepackage{wrapfig}
\usepackage{filecontents}
\usepackage{enumitem}
\usepackage{wrapfig}
\usepackage{dblfloatfix}
\usepackage{pbalance}

\begin{document}

\date{}

\title[CAMEO: A Causal Transfer Learning Approach for Performance Optimization]{CAMEO: A Causal Transfer Learning Approach for Performance Optimization of Configurable Computer Systems}

\author{Md Shahriar Iqbal}
\affiliation{%
  \institution{University of South Carolina}
  \country{}
}

\author{Ziyuan Zhong}
\affiliation{%
  \institution{Columbia University}
  \country{}
}

\author{Iftakhar Ahmad}
\affiliation{%
  \institution{University of South Carolina}
  \country{}
}

\author{Baishakhi Ray}
\affiliation{%
  \institution{Columbia University}
  \country{}
}

\author{Pooyan Jamshidi}
\affiliation{%
  \institution{University of South Carolina}
  \country{}
}

\begin{abstract}
Modern computer systems are highly configurable, with hundreds of configuration options that interact, resulting in an enormous configuration space. As a result, optimizing performance goals (e.g., latency) in such systems is challenging due to frequent uncertainties in their environments (e.g., workload fluctuations). Recently, transfer learning has been applied to address this problem by reusing knowledge from configuration measurements from the source environments, where it is cheaper to intervene than the target environment, where any intervention is costly or impossible.  Recent empirical research showed that statistical models can perform poorly when the deployment environment changes because the behavior of certain variables in the models can change dramatically from source to target. To address this issue, we propose \ourapproach---a method that identifies invariant causal predictors under environmental changes, allowing the optimization process to operate in a reduced search space, leading to faster optimization of system performance. We demonstrate significant performance improvements over state-of-the-art optimization methods in \textsc{MLperf} deep learning systems, a video analytics pipeline, and a database system. 

\end{abstract}



\maketitle


\section{Introduction}
\label{sec:intro}
Modern computer systems are continuously deployed in heterogeneous environments (e.g., cloud, FPGA, SoC) and are highly configurable across the software/hardware stack~\cite{jamshidi2014autonomic,pahl2018architectural}.
In such highly configurable systems, optimizing performance indicators, e.g., latency and energy, is crucial for faster data processing, better user satisfaction, and lower application maintenance cost~\cite{wang2021morphling,ding2022cello}. One possible way to achieve these goals is to tune the systems with configuration options across the stack, such as \textit{cpu frequency}, \textit{swappiness}, and \textit{memory growth}, to achieve optimal performance~\cite{xu2015hey,blocher2021switches,colin2018reconfigurable}.

\begin{wrapfigure}{r}{0.5\columnwidth}
\includegraphics[width=0.5\columnwidth]{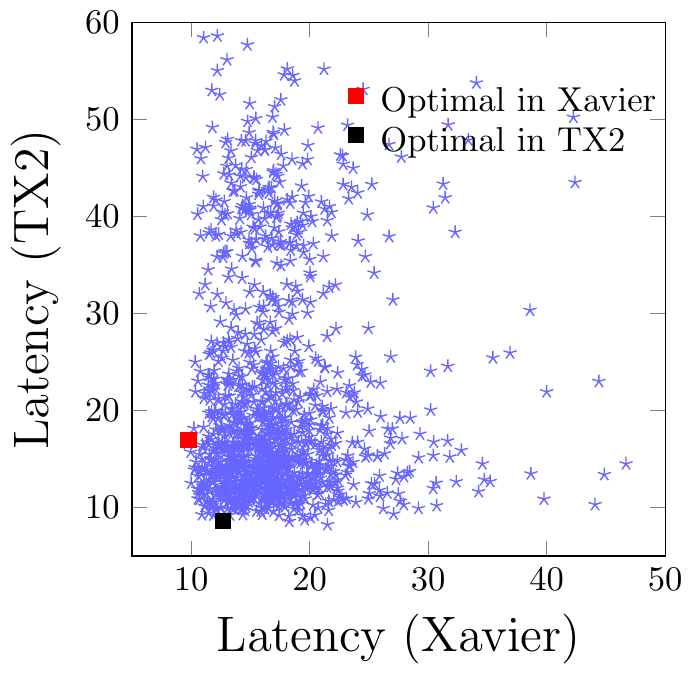}
\vspace{-8mm}
\caption{\small{The optimal configuration for \textsc{MLPerf Object Detection} pipeline deployed on TX2 is not optimal in Xavier.
}}
\label{fig:hw_change_intro}
\end{wrapfigure}
Finding an optimal configuration in a highly configurable system, however, is challenging~\cite{JC:MASCOTS16,wang2018understanding,halin2019test,velez2022study,acher2019learning,chen2023performance}: (i) Each component in the system stack, i.e., software, hardware, OS, etc., has many configuration options that interact with each other, giving rise to combinatorial configuration space, (ii) estimating the effect of configurations on performance is expensive as one needs to collect run-time behavior of the system for each configuration, and (iii) unknown constraints exist among configuration options, giving rise to many invalid configurations. Moreover, to meet growing user requirements and reduce service management costs, underlying systems often undergo environmental changes, that is, hardware updates, changes in deployment topology, etc.~\cite{ding2021generalizable}. Therefore, optimizing the performance of these evolving systems becomes even more challenging since there is no guarantee that the optimal configurations found in one environment will remain optimal in a different environment~\cite{JSVKPA:ASE17,JVKS:FSE18,JC:MASCOTS16}\footnote{we define an environment as a combination of hardware, workload, software, and deployment topology} as shown in Figure~\ref{fig:hw_change_intro}. \looseness=-1



To address these challenges, in real-world deployment scenarios, developers often use a staging (development) environment, a miniature of a production environment, for testing and debugging. Developers collect many experimentation and performance evaluations in staging environments (hereafter, we call them source environments) to understand the performance behavior of the system (what configurations potentially produce performance anomalies, what configurations produce stable performance, or where good configurations lie). Developers then use that knowledge in target production settings for downstream performance optimizations or debugging. However, in most cases, the staging environment result is completely different from the production result, resulting in a misleading or even wrong indication about the configurations that produce optimal performance. These differences in the results occur mainly due to the hardware gap or workload differences between the development environment and the production environment. For example, the workload of an ML system may surge, and as a result, the batch size behind the model server needs to increase to sustain the latency requirement; however, due to the different memory hierarchy and CPU cores between the source and the target environments, the optimal setting for inter-op parallelism of the model server would be vastly different in each environment \cite{salmani2023reconciling}.

\noindent\textbf{Existing works and gap.} \textit{Performance optimization in configurable systems.} Several approaches have been proposed for performance optimization of configurable systems, e.g. Bayesian optimization (BO)~\cite{hutter2011sequential,yigitbasi2013towards,wu2015deep,alipourfard2017cherrypick,menon2020auto,JC:MASCOTS16,JVKS:FSE18}, BO with regression~\cite{ding2022cello}, prediction models~\cite{chen2021efficient}, search space modification~\cite{hsu2018scout}, online few shot learning~\cite{blocher2021switches}, and uniform random sampling and random search algorithms~\cite{oh2022finding}. However, using these approaches in a production environment requires many queries, which are often too expensive to collect or may be infeasible to perform. 
The optimal configuration found by these methods in a source environment is also suboptimal for the targets, as the optimal configuration determined in the source environment usually no longer remains optimal in the other (see \Cref{fig:hw_change_intro} for an example).

\noindent\textit{Transfer Learning for Performance Analysis.} In real-world deployment scenarios, developers typically have access to performance evaluations of different configurations from a staging environment. Exploiting this additional information using transfer learning can result in efficient optimization, as demonstrated by recent work ~\cite{JVKS:FSE18,krishna2019whence,iqbal_transfer_2019,martin2021transfer,lesoil2022transferring,JVKSK:SEAMS17}. 
For example, searching for optimized performance in the target setting can use the summary statistics of the models built using the performance of the source~\cite{zhang2021restune}. 
However, each environmental change can potentially cause a distribution shift. The ML models used in these transfer learning methods are vulnerable to spurious correlations, which do not hold between distribution shifts and result in inferior performance ~\cite{zhou2021examining,ming2022impact,iqbal2022unicorn} (see \Cref{sec:motivation_causal} for an example).

\begin{table}
\vspace{+2mm}
\caption{\small Comparison of \ourapproach with state-of-the-art system performance optimization approaches. }
    \centering
\resizebox{\columnwidth}{!}{
\begin{tabular}{ccccccc}

\hline
    \centering
    Feature&\smac&\cello& \unicorn& \restune&\restuner&\ourapproach \\

    \hline
    Detects Spurious Features&\ding{55}&\ding{55}&\ding{51}&\ding{55}&\ding{55}&\cellcolor{red!20}\ding{51}\\
    Handles Distribution Shift&\ding{55}&\ding{55}&\ding{55}&\ding{55}&\ding{51}&\cellcolor{red!20}\ding{51}\\
    Suitable for Benchmarks&\ding{51}&\ding{51}&\ding{51}&\ding{55}&\ding{51}&\cellcolor{red!20}\ding{51}\\
Knowledge Reuse&\ding{55}&\ding{55}&\ding{55}&\ding{55}&\ding{51}&\cellcolor{red!20}\ding{51}\\
Constrained
Optimization&\ding{55}&\ding{51}&\ding{55}&\ding{51}&\ding{51}&\cellcolor{red!20}\ding{51}\\
    \hline
\end{tabular}}
\label{tab:sota_vs_cameo}
\end{table}

\begin{figure}[t]
    \centering
    \includegraphics[width=\columnwidth]{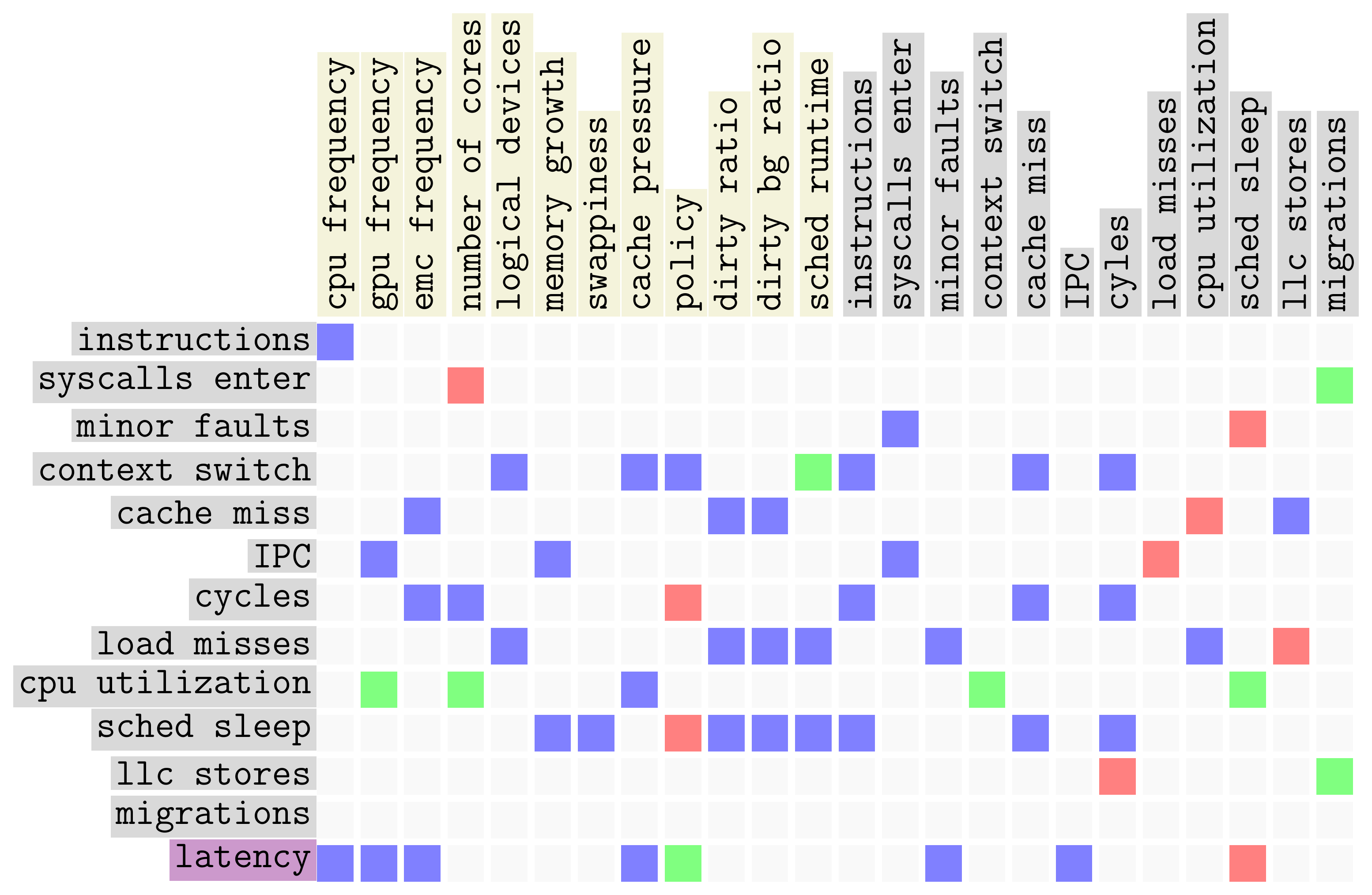}
    \caption{\small{There is a significant overlap between the causal structures (the common edges are represented as blue squares) developed in different environments (e.g., Jetson TX2 and Xavier). Some edges unique to the source (green squares) or target (red squares) also exist.}}
    \label{fig:graph_diff}
\end{figure}

\noindent\textit{Usage of Causal Analysis in Configurable Systems.} To address the problem of spurious correlations, recent work has leveraged causal inference \cite{iqbal2022unicorn, dubslaff2022causality,siegmund2022green} to build a causal performance model\footnote{A causal performance model is an acyclic-directed mixed graph, with nodes being variables and arrows being causal connections. It represents the dependencies (a.k.a. causal structures) between configuration options, system events, and performance objectives.} that captures the dependencies among configuration options, system events, and performance objectives. However, the causal graphs in the source and target can still have some differences (see \Cref{fig:graph_diff} for an example). Recent work~\cite{iqbal2022unicorn} shows that the source causal model could be reused for performance debugging in the target environment; however, further measurements are needed for the learning and optimization of the performance model. 

In summary, all these existing works are suboptimal for performance optimization when the environment changes because the knowledge extracted by these methods from the source (i.e., optimal configuration) has changed and cannot be directly applied to the target, the model (i.e., ML-based transfer learning model) may capture spurious correlations, or the model (i.e., causal model) is mostly stable but needs further adaptation in the target environment (see Table~\ref{tab:sota_vs_cameo}).

\noindent\textbf{Our approach.}
An ideal optimization approach should leverage the knowledge derived from the source, which is a close replica of the target environment with a cheaper experimentation cost. Our key insight is that by using causal reasoning, we should be able to identify the non-spurious invariances across environments that truly impact the performance behavior of the system. These invariances can then be transferred to the target environment for performance optimization tasks, thus reducing the need for observational data in the production environment. Therefore, we will reduce the cost of optimization tasks without compromising accuracy. 

\begin{figure}[t]
    \centering
    \subfloat[source (TX2)]{
	\begin{minipage}[c][1\width]{
	   0.32\linewidth}
	   \centering
	   \includegraphics[width=\textwidth]{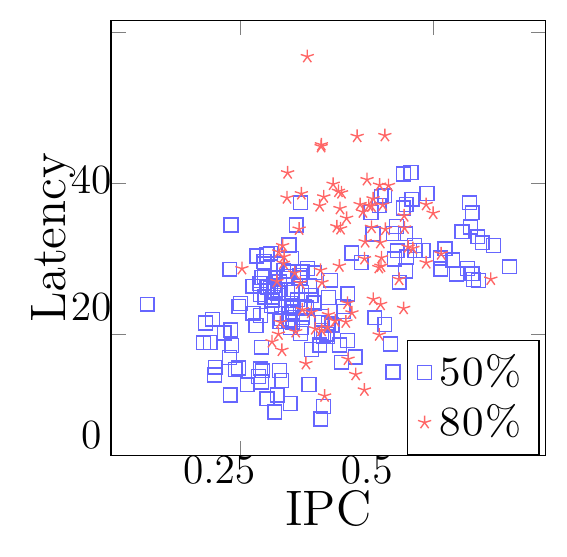}
	\end{minipage}}
 \hfill
    \subfloat[target (Xavier)]{
	\begin{minipage}[c][1\width]{
	   0.32\linewidth}
	   \centering
	   \includegraphics[width=\textwidth]{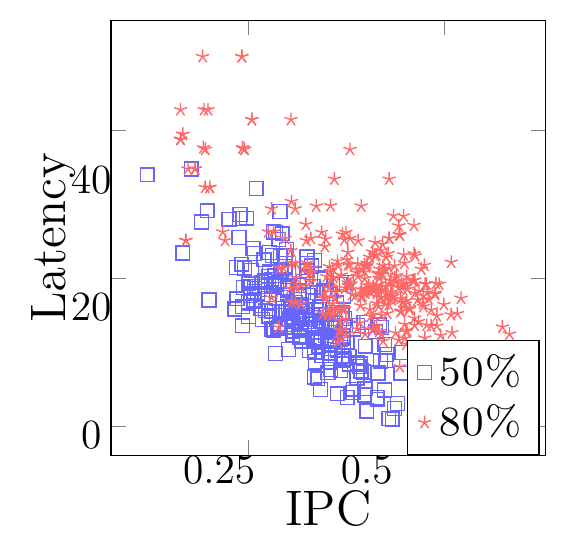}
	\end{minipage}}
 \hfill 
    \subfloat[true relationship]{
	\begin{minipage}[c][1\width]{
	   0.32\linewidth}
	   \centering
	   \includegraphics[width=\textwidth]{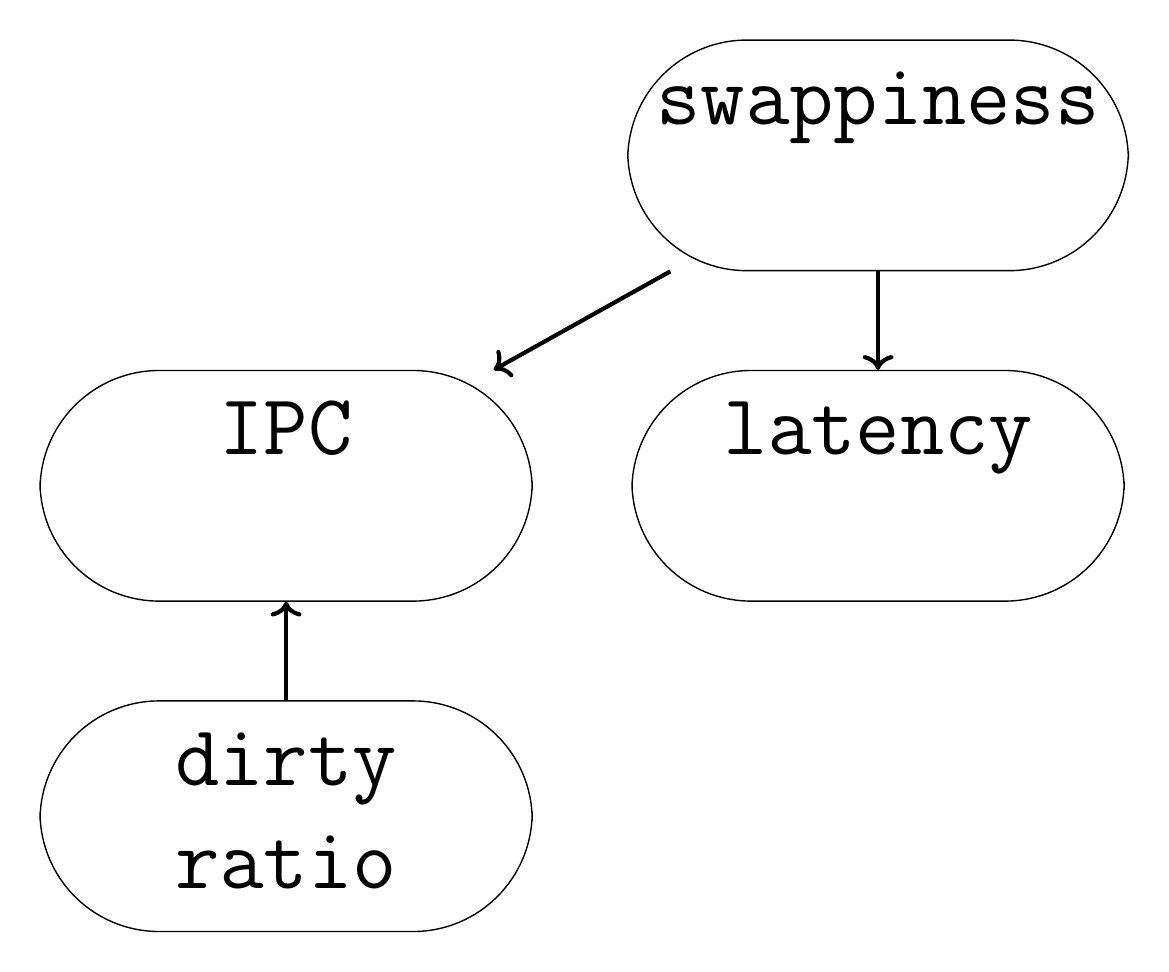}
	\end{minipage}}

    \caption{\small{(a)-(b) The relationship between \textsf{IPC} and \textsf{latency} reverse from source (TX2) to target (\xavier) while the relationship between \textsf{swappiness} (the values are denoted as colors) and \textsf{latency} stays invariant. (c) The true causal relationship among the relevant variables.}}
    \label{fig:spurious}
\end{figure}

To this end, we propose \ourapproach (\textbf{Ca}usal \textbf{M}ulti \textbf{E}nvironment \textbf{O}ptimization), a causal transfer-based optimization algorithm aimed at overcoming
the limitation of prior approaches. Our approach is built on top of two previous works, JUMBO (a multitask BO method) \cite{hakhamaneshi2021jumbo} and CBO (a causal BO method) \cite{causalBO}. A typical BO approach consists of two main elements: the surrogate model and the acquisition function. The surrogate model tries to predict the performance objective when given a configuration, and the acquisition function assigns a score to each configuration and chooses the one with the highest score to query for the next iteration.
In \ourapproach, we first build two causal performance models to learn the dependency among the configuration options, system events, and performance objectives for each environment using the previous performance measurements of the source environment and a considerably smaller number of measurements of the target environment. After that, we simultaneously train two Causal Gaussian Processes (CGPs) (which leverage the causal performance models when estimating means and variances) as two surrogate models: a warm CGP in the source and a cold CGP in the target.
The acquisition function combines the individual acquisition functions of both CGPs to leverage knowledge from both the source and target. This way of combining individual acquisition functions of both CGPs allows one to rely only on the core features from the source environment that remain stable across environments and update belief about the environment-specific features in the target, making the optimization more effective. 

\noindent\textbf{Evaluation.}~
We evaluated \ourapproach in terms of its \emph{effectiveness, sensitivity}, and \emph{scalability}, and compared it with four state-of-the-art performance optimization techniques (\smac~\cite{hutter2011sequential}, \restune and \restuner~\cite{zhang2021restune}, \cello~\cite{ding2022cello}, and \unicorn~\cite{iqbal2022unicorn}) using five real-world highly configurable systems, including three \textsc{MLperf} pipelines (object detection, natural language processing and speech recognition), a video analytics pipeline, and a database system, deployed on edge and cloud under different environmental changes. Our results indicate that \ourapproach improves latency by 3.7$\times$ and energy by 5.6$\times$ on average than the best baseline optimization approach, \restuner.

\noindent \textbf{Contributions.}
Our contributions are as follows: 
\begin{itemize}[noitemsep, nolistsep, leftmargin=*]
    \item We propose \ourapproach, a novel causal transfer-based approach that allows faster optimization of software systems when the environment changes. 
    \ourapproach is one of the first approaches to use causal transfer learning for the optimization of the performance of configurable systems.
    \item We conducted a comprehensive evaluation of \ourapproach by comparing it with state-of-the-art optimization methods in five highly configurable systems in the real world under a range of different environmental changes and studied the effectiveness of design explorations with different varieties and severity of environmental changes and showed the scalability of our approach to colossal configuration spaces. The artifacts and supplementary materials can be found at \href{https://github.com/softsys4ai/CAMEO}{\color{blue!80} https://github.com/softsys4ai/CAMEO}.
\end{itemize}

\


\section{Motivation and Insights}
\label{sec:motivation}

In this section, we motivate our approach by illustrating why causal reasoning can contribute to more effective optimization of system performance. In particular, we focus on how the properties of the causal performance models can be leveraged across environments. For this purpose, we used the \textsc{Mlperf Object Detection}~\cite{reddi2020mlperf} pipeline as part of the MLPerf Inference Benchmark\footnote{\url{https://mlcommons.org/en/inference-edge-30/}} following the benchmark rules\footnote{\url{https://github.com/mlcommons/inference\_policies/blob/master/inference\_rules.adoc}}, with the following setup: Model: Resnet50-v1.5; Test scenario: Offline; Metric: inference latency; Workload: 5000 ImageNet samples; workload generator: Mlperf Load Generator; Source Hardware: Jetson TX2; Target hardware: Jetson Xavier and TX1.
For better control, we limit the configuration space to 28 options across the stack---4 hardware options (e.g., \textsf{cpu cores}),  22 OS options (e.g., \textsf{dirty ratio}), and 2 compiler options (e.g., \textsf{allow memory}). 
We sampled 2,000 random configurations and measured the inference latency in each environment. We also collected performance counters and system events statistics using \textit{Linux perf profiler}\footnote{\url{https://perf.wiki.kernel.org/}}.


\subsection{Why performance optimization using causal reasoning is more effective?}
\label{sec:motivation_causal}
To deploy a configurable computer system such as \textsc{Mlperf Object Detection} in a new environment with low latency and energy consumption, the dominant approach is to train a performance model using a limited number of samples and use the model to predict performance for unmeasured configurations and select the configuration with the optimal performance.  
To show how spurious features could mislead performance optimization, we investigate the impact of confounders and how they make it difficult for an ML model to determine the accurate relationship between configuration options and performance objectives. We perform a sandbox experiment where we carefully tune \textsf{swappiness}~\footnote{\textsf{swappiness} is the rate at which the kernel moves pages into and out of the physical memory. The higher the value, the more aggressive the kernel will be in moving the pages out of physical memory to the swap memory.} and \textsf{dirty ratio}~\footnote{\textsf{dirty ratio} is the value that represents the percentage of physical memory that can consume dirty pages before all processes must write dirty buffers back to the disk.} both in source and target, while leaving all other options at their default values. Here, the observational data collected from the experiment indicates that as \textsf{IPC}~\footnote{\textsf{IPC} represents instruction per cycle, which is the average number of instructions executed for each clock cycle.} (one of the system events) increases, \textsf{latency} increases, which is a spurious proportional relationship.
Relying on spurious features (\textsf{IPC} in this example) can lead to poor performance predictions (as one might try to reduce \textsf{IPC} and expect lower \textsf{latency} but end up getting higher \textsf{latency}) when the environment changes because they are susceptible to \textit{correlation shifts}---i.e., the direction of correlation may change across environments. As shown in Figure~\ref{fig:spurious}(a)-(b), a correlation shift occurs in this sandbox experiment, as \textsf{IPC} is positively correlated with \textsf{latency} in the source, but negatively correlated in the target. 

To investigate the reason behind the correlation shift, we group the data based on their \textsf{swappiness} (50\% and 80\%, respectively) and observe that the correlation between \textsf{swappiness} and \textsf{latency} remains the same (larger swappiness implies higher latency in both environments) whereas the correlation between \textsf{swappiness} and \textsf{IPC} reverses (from proportional to inverse proportional) as shown in Figure~\ref{fig:spurious}(a)-(b). Figure~\ref{fig:spurious}(c) shows the causal structure where \textsf{swappiness} is a \textit{common cause} of both \textsf{IPC} and \textsf{latency}. \textsf{swappiness} should be considered for \textsf{latency} since it remains invariant across environments. On the contrary, the relationship between \textsf{IPC} and \textsf{latency} is environment dependent, and their correlation can change when another confounder variable, \textsf{dirty ratio}, is different in source and target. In our example, since the source has 4$\times$ lower physical memory than the target, the allocated memory for the dirty pages becomes filled sooner and must be returned to the disk. As a result, the source will have higher \textsf{IPC} for a lower value of \textsf{swappiness} as the dirty pages will be flushed before the limit for \textsf{swappiness} is reached. However, the application is not making any forward progress here, resulting in increased \textsf{latency}. In the target (due to larger memory), the dirty pages might never become full, and only \textsf{swappiness} would cause the \textsf{IPC} to be positively correlated to \textsf{latency}. 
The example in \Cref{fig:spurious} shows that the casual model can better capture the \textit{data generation process} as it only relies on invariant causal mechanisms (\textsf{swappiness} for \textsf{latency}) and can remove spurious correlations (\textsf{IPC} for \textsf{latency}) that are specific to a particular environment. Therefore, causal models may suffice to predict the consequences of interventions (\textit{what if} scenarios) on variables to particular values for effective search during optimization and allow better explorations in limited budget scenarios. 

\begin{table}[]
\caption{\small{ML-based regressors (GPR, RFR) have higher generalization error compared to causal-based regressor (CGPR).}}
    \centering
\begin{tabular}{cccccc}

\hline
    \centering
    Source&Target&KL Div.&\multicolumn{3}{c}{Prediction Error (\%)}\\
    \cline{4-6}
    &&&GPR&RFR&CGPR\\
    \hline
    TX2&Xavier&476&22.4&25.6&\cellcolor{red!20}{11.2}\\
    TX2&TX1&519&27.6&23.2&\cellcolor{red!20}{11.4}\\
    \hline
\end{tabular}

\label{tab:ood_regr}
\end{table}

To show the benefits of correctly identifying the invariant features, we train different ML-based regressors, e.g., the Gaussian Process Regressor (GPR) and the Random Forest Regressor (RFR), using data collected for the sandbox system deployed in TX2 and determined their prediction error in TX1 and \textsc{Xavier} (shown in Table~\ref{tab:ood_regr}). Here, we observe that the ML-based regressors have considerably higher errors in the target environment despite low source errors. The prediction error increases further as the distributions become more dissimilar (indicated by a higher KL-divergence value). In contrast, the causal approach, Causal Gaussian Process Regressor (CGPR), has a considerably lower error and remains stable as the degree of distribution shift increases. 

\begin{tcolorbox}[colback=blue!5!white,colframe=blue!75!black]
\noindent\textbf{Takeaway 1} Causal models generalize better in performance prediction tasks across environments by distinguishing invariant from spurious features.
\end{tcolorbox}

\subsection{Learning from Causal Structural Properties in Various Environments}
\label{sec:motivation_causal_in_various_environments}
As we have established that a causal model can be reliably used for performance predictions in new environments, we next study the properties of the causal graph that can be exploited for faster optimization. We build a causal graph using a causal structure discovery algorithm~\cite{spirtes2000causation} in the source and target, respectively, and compare them. As shown in \Cref{fig:graph_diff}, both causal graphs are sparse (the white squares indicate no dependency relationship exists) and share a significant overlap (the blue squares indicate the edges present in both). Therefore, a causal model developed in one environment can be leveraged in another as prior knowledge. However, reusing the causal graph entirely might induce some wrong biases as the causal graphs in the two environments are not identical (the green and red squares indicate the edges present uniquely in the source and target, respectively). We must discover the new causal connections (indicated by the red squares) based on the observation. Since the number of edges that must be discovered is small, this can be easily done with a small number of observational samples from the target environment.

\begin{figure}[t]
    \centering
    \includegraphics[width=\columnwidth]{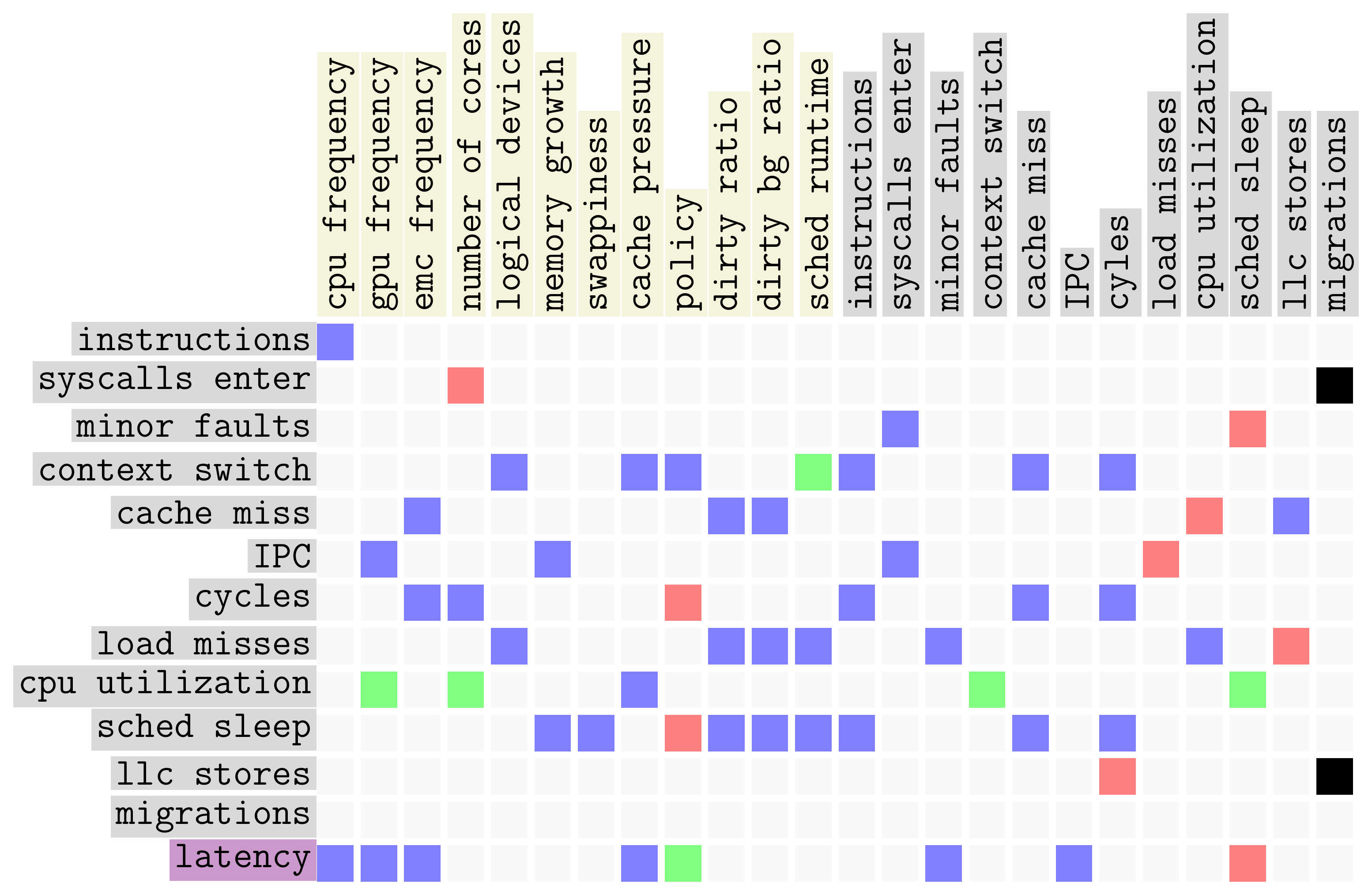}
    \caption{\small{Combining the top K nodes' Markov blankets eliminates the wrong biases (shown as black squares). }}
    \label{fig:topk}
\vspace{-1mm}
\end{figure}


To eliminate biases, we need to remove unique edges of the source. Removal operations can be accomplished by performing interventions that estimate the effects of deliberate actions. 
For example, we measure how the distribution of an outcome (e.g., \textsf{latency} $\mathcal{Y}$) would change if we intervened during the data collection process by forcing the variable \textsf{cpu frequency} $\mathcal{O}_{i}$ to a certain value $o_i$ while retaining the other variables as is. 
We can estimate the outcome of the intervention by modifying the causal performance model to reflect our intervention and applying Pearl's \emph{do-calculus}~\cite{pearl2009causality}, which is denoted by $Pr(\mathcal{Y}~|~do(\mathcal{O}_{i}= o_i))$. However, since many configurations need to be measured, it is not feasible to perform interventions to estimate the existence of every edge. Instead, we can significantly reduce the number of configurations by avoiding interventions on nodes with limited causal effects on the performance objective. For this purpose, we rank the causal effects of all existing nodes on \textsf{latency} 

and observe that only one source-specific edge (\textsf{policy}) is among the top 10 most influential nodes. 
Therefore, we can select the \textit{ K} nodes with the highest causal effects and combine the \textit{Markov blanket}~\footnote{A Markov blanket of a node includes all its parents, children, and children's parents.} of them, which would eliminate all the nodes that have lower causal effects. 
\begin{wrapfigure}{r}{0.5\columnwidth}
    \centering
    
    \includegraphics[width=0.5\columnwidth]{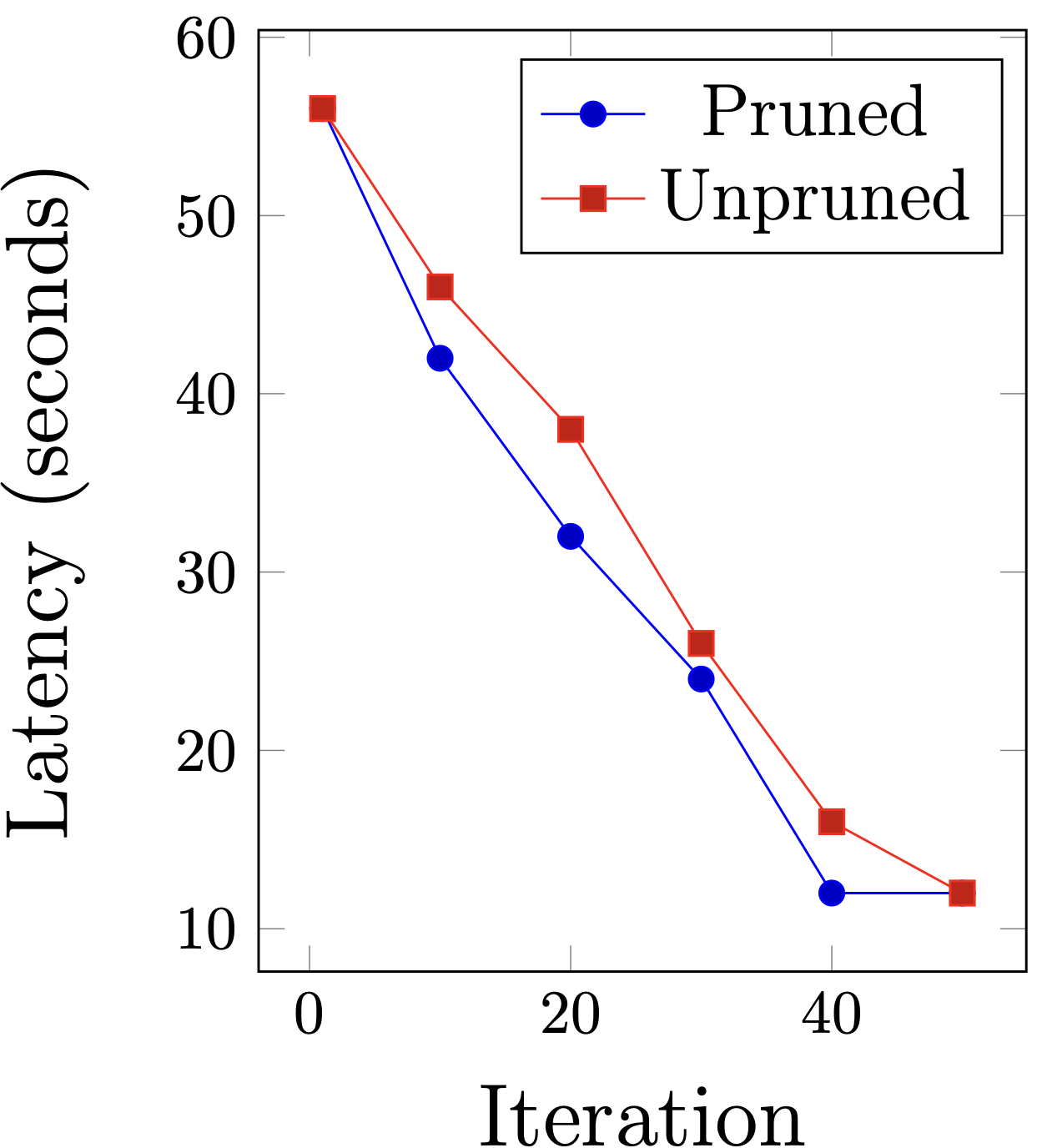}
    \vspace{-5mm}
    \caption{\small{ Pruning edges with a Markov blanket identifies the optimal configuration faster. }}
    \label{fig:topk}
\vspace{-1mm}
\end{wrapfigure}
In our example, 
if we select K=6 with Markov blankets then the wrong biases, \textsf{migrations}->\textsf{syscalls enter} and \textsf{migrations}->\textsf{llc stores} (the nodes marked by black in Figure~\ref{fig:topk}(b)), are eliminated. 
Figure~\ref{fig:topk}(a) shows that pruning the edges helps to reach the optimal value 19\% faster. 
Therefore, \textit{we require an approach that relies on intervening only in the top K nodes based on the source knowledge in the target environment}.


\begin{tcolorbox}[colback=blue!5!white,colframe=blue!75!black]
\noindent\textbf{Takeaway 2} Employing rich knowledge in a causal performance model, we can intervene in specific configurations to learn the most about the underlying causal structure and be able to gather the most relevant data under a limited budget.
\end{tcolorbox}

\section{\ourapproach Design}
\label{methodology}

In this section, we present \ourapproach---a framework for performance optimization of highly configurable systems. 

\begin{figure*}[t]
\centering
\includegraphics[width=\textwidth]{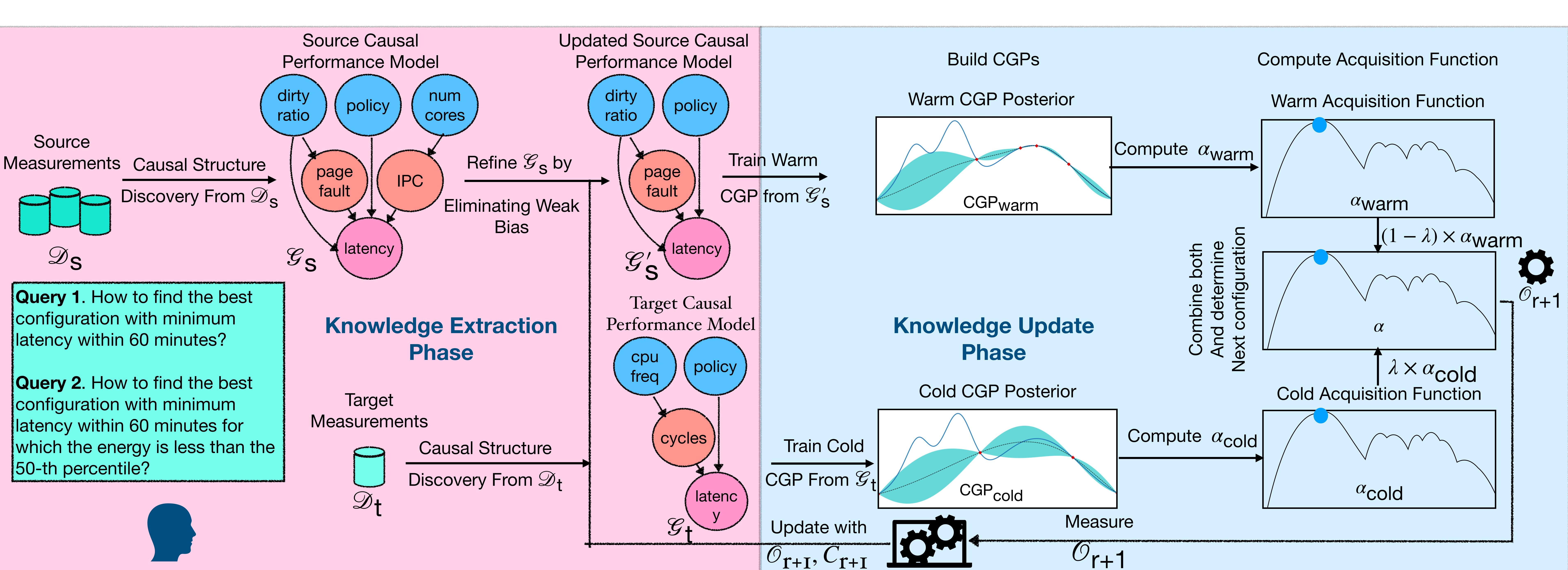}
\caption{\small{Overview of \ourapproach}.}
\label{fig:overview}
\end{figure*}

\subsection{Problem Formulation}

Let us consider a highly configurable system of interest with configuration space ${\mathcal{O}}$, system events and performance counters space $\mathcal{C}$, and a performance objective  $\mathcal{Y}$. Denote $\mathcal{O}_i$ to be the $i^{th}$ configuration option of a system, which can be set to a range of different values (e.g., categorical, Boolean, and numerical). The configuration space
is a Cartesian product of all hardware, software, and application-specific options: $\mathcal{O} = Domain(\mathcal{O}_1) \times ... \times Domain(\mathcal{O}_d$), where d is the number of options.
Configuration options and system events are jointly represented as a vector 
$\mathcal{X}=(\mathcal{O},\mathcal{C})$.
We assume that in each environment $e \in \mathcal{E}$ (a combination of hardware, workload, software, and deployment topology), the variables $(\mathcal{X}_e,\mathcal{Y}_e)$ have a joint distribution $\mathcal{P}_e$. In the source environment $e_{\text{s}}$, there are $n$ independent and identically distributed (i.i.d) 
observations. The task is to find a near-optimal configuration, $o^*$, with a fixed measurement budget, $\mathcal{\beta}$, in the target environment, $e_{\text{t}}$, that results in Pareto-optimal performance: 
\begin{equation}
   o^* = \text{argmin}_{o \in \mathcal{O}} \mathcal{Y}_{{e}_\text{t}}(o) \ \text{within} \ \beta,
\end{equation}
where $\mathcal{O}$ represents the configuration space, $\mathcal{Y}$ is a set of performance metrics measured in the target environment $e_{\text{t}}$.

\subsection{\ourapproach Overview}





\ourapproach is a causal transfer learning optimization algorithm that enables developers and users of highly configurable computer systems to optimize performance objectives such as latency, energy, and throughput when the deployment environment changes. Figure~\ref{fig:overview} illustrates the overall design of our approach. \ourapproach works in two phases: (i) \textit{knowledge extraction phase}, and (ii) \textit{knowledge update phase}. In the knowledge extraction phase, \ourapproach first determines the user requirements using a query engine. Then, it learns a causal performance model $\mathcal{G}_{\text{s}}$ using cheaper offline performance measurements $\mathcal{D}_{\text{s}}$ from the source environment $e_{\text{s}}$, which is later reused to obtain meaningful information that is shared with the target environment $e_{\text{t}}$ for faster optimization. As performance evaluations in the target are expensive, this way of warm-starting the optimization process by reusing the causal performance model $\mathcal{G}_\text{s}$ enables us to navigate the configuration space more effectively with less number of interventions in the target. However, relying solely on the source's information is insufficient to effectively optimize performance in the target due to the differences across environments (as shown in \Cref{sec:motivation_causal_in_various_environments}). Therefore, in the knowledge update phase, \ourapproach employs an active learning mechanism combining the source causal performance model $\mathcal{G}_{\text{s}}$ with a new causal performance model $\mathcal{G}_{\text{t}}$ collected from a small number of samples, $\mathcal{D}_{\text{t}}$, from the target environment. 

Once the two causal performance models are constructed, we simultaneously train two causal Gaussian processes (CGPs) as the surrogate models---$\text{CGP}_\text{warm}$ and $\text{CGP}_\text{cold}$---to model performance objective $\mathcal{Y}$ from $\mathcal{G}_{\text{s}}$ and $\mathcal{G}_{\text{t}}$, respectively. The two CGPs operate on different input spaces. $\text{CGP}_\text{warm}$ works on a reduced configuration space that is derived from $\mathcal{G}_{\text{s}}$. In contrast, to ensure that any information omitted in the source is not left undiscovered in the target, $\text{CGP}_\text{cold}$ works on the entire configuration space. We integrate the posterior estimates from both $\text{CGP}_\text{warm}$ and $\text{CGP}_\text{cold}$ to develop an acquisition function $\alpha$ that can regulate the information from two CGPs through a controlling variable $\lambda$. The larger $\lambda$, the more we rely on the information in $\text{CGP}_\text{warm}$.
Next, we evaluate our acquisition function $\alpha$ for different configurations and select the one for which the $\alpha$ value is maximum for observation or intervention. The choice of observation and intervention for performance evaluation is guided by an exploration coefficient $\epsilon$. Finally, we use the newly evaluated configurations to update the causal performance and surrogate models. We continue the active learning loop until the stopping criterion is met (i.e., the maximum budget $\beta$ is exhausted or convergence is achieved). The pseudocode for our approach is provided in \Cref{alg:cameo}. 
\begin{figure}[t]
\centering
\includegraphics[width=\columnwidth]{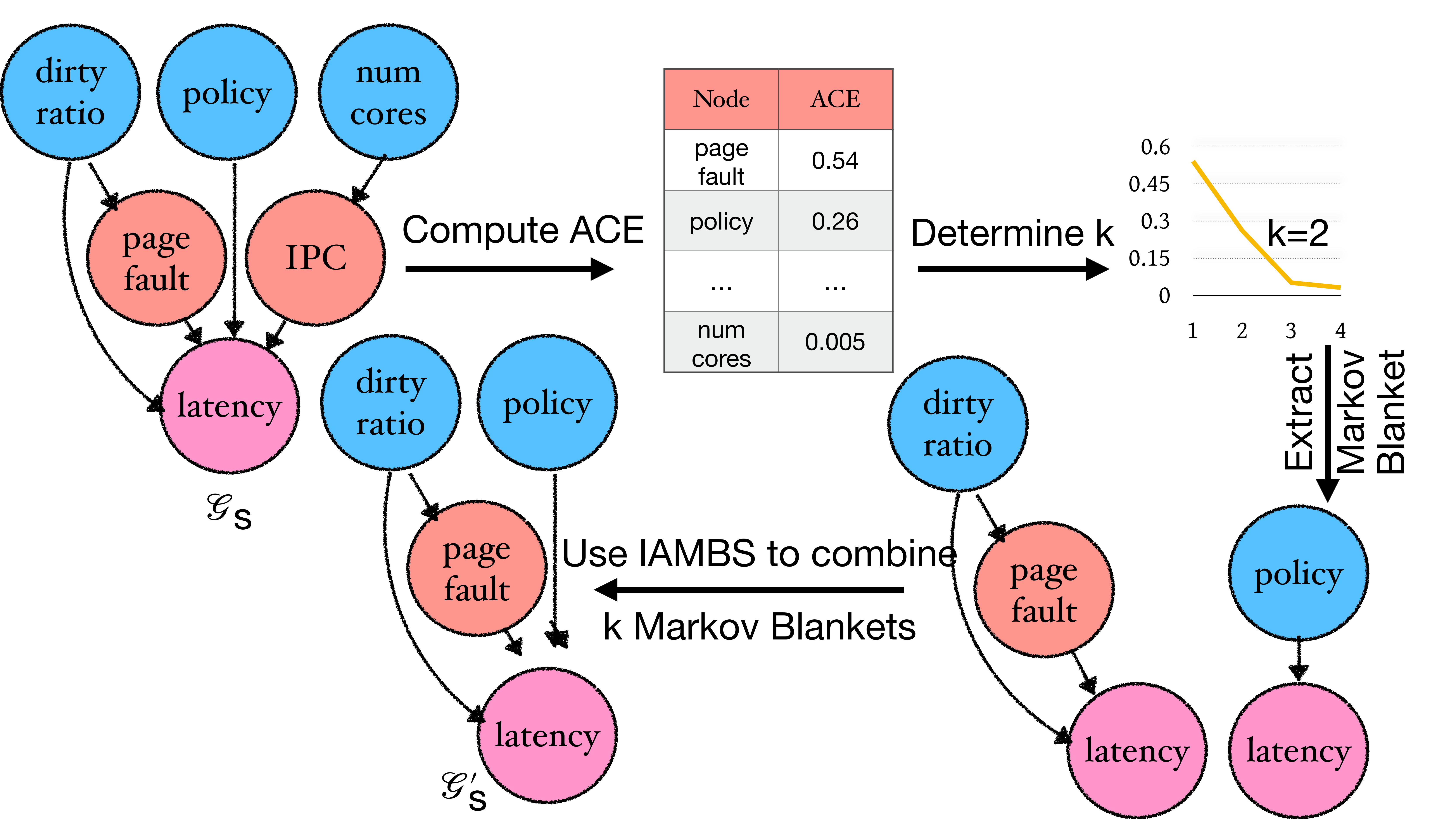}
\caption{\small{Refining the causal performance model from the source to eliminate unwanted information.}}
\label{fig:knowledge_extract}
\end{figure}

\subsection{Knowledge Extraction Phase}
We next describe the offline knowledge extraction phase. 

\smallskip
\noindent
\textbf{User query translation.}
A developer can use \ourapproach to find the optimal configurations that optimize a system's performance objectives in a target environment within a limited experimentation budget $\beta$. The developer can start the optimization process by querying \ourapproach with requests like \textit{"How to improve latency within 1 hour or 50 samples"} or \textit{"I want to find the configuration with minimum energy for which latency is less than 20 seconds within 45 minutes?"}. The query engine initially translates user requests to determine the allowable budget $\beta$, constraints $\psi$, and the performance goal $\mathcal{Y}$ to optimize. In the first query, the budget is 1 hour or 50 samples, the performance objective is latency, and no constraints exist. In the second query, the budget is 45 minutes, the performance objective is energy, and the constraint is a latency of less than 20 seconds. The query translator extracts this information by directly accepting user inputs with some fixed guided keyword directives.

\smallskip
\noindent
\textbf{Learning causal performance model.}
We begin by building two causal performance models: $\mathcal{G}_{\text{s}}$ and $\mathcal{G}_{\text{t}}$ using the offline performance evaluation dataset $\mathcal{D}_{\text{s}}$ from the source with $n$ configurations and the performance dataset $\mathcal{D}_{\text{t}}$ from the target with randomly sampled $m$ initial configurations, respectively. We use an existing structure discovery algorithm \emph{fast causal inference} (FCI) to learn $\mathcal{G}_{\text{s}}$ and $\mathcal{G}_{\text{t}}$ that describes the causal relations among configuration options $\mathcal{O}_i$, system events and performance counters $\mathcal{C}_i$, and performance objectives $\mathcal{Y}$. 
We select FCI as the causal structure discovery algorithm because (i) it accommodates variables that belong to various data types such as nominal, ordinal, and categorical data common across the system stack, and (ii) it accommodates the existence of unobserved confounders~\cite{spirtes2000causation,ogarrio2016hybrid, glymour2019review}. This is crucial because we do not assume absolute knowledge of configuration space, so there may be configurations in which we cannot intervene or system events we have not observed.
FCI operates in three stages. First, we construct a fully connected undirected graph where each variable is connected to every other variable. Second, we use statistical independence tests (Fisher's z test for continuous variables and mutual information for discrete variables) to remove edges between independent variables. Finally, we orient undirected edges using prescribed edge orientation
rules~\cite{spirtes2000causation,ogarrio2016hybrid,glymour2019review,colombo2012learning,colombo2014order} to produce a \textit{partial ancestral graph} (or PAG). In addition to both directed and undirected edges, a PAG also contains partially directed edges that need to be resolved to generate an acyclic-directed mixed graph (ADMG), i.e., we must fully orient partially directed edges with the correct edge orientation. This work uses an information-theoretic approach to automatically orient partially directed edges using the \emph{LatentSearch} algorithm~\cite{Kocaoglu2017} by entropic causal discovery.

\smallskip
\noindent
\textbf{Refining causal performance model.}
Now that we have constructed the causal performance models based on the invariant features, we may be tempted to directly reuse the source model $\mathcal{G}_{\text{s}}$ in the target to warm-start the optimization process. However, since some edges are specific to the source (as discussed in \Cref{sec:motivation_causal_in_various_environments}), directly reusing $\mathcal{G}_{\text{s}}$ will bias the optimization in the target. To avoid wasting the budget allocated for the online optimization procedure, we try to minimize those biases as much as possible in this offline phase. To do so, we transfer the Markov blanket (Mb) of the top $k$ nodes ranked according to their causal effects on the performance objective to eliminate unwanted information. 
Higher causal effects indicate a stronger influence of the configuration option on performance. When scaling option values within a constant context, options with higher causal effects become top features. This is an important step, as we need to rely on the optimal core features that remain invariant when a performance distribution shift happens to reason better in the new environment. 
Theoretically, the Mb of a node is the best solution to the feature selection problem for that node~\cite{javidian2021scalable}. The variables in the Mb can be confidently employed as causally informative features in the target because they provide a thorough picture of the local causal structure around the variable. Initially, we determine $k$ using the method proposed in~\cite{hamerly2003learning}. Then we extract the Mb of the $k$ nodes to determine the final $\mathcal{G}_\text{s}$ that will be reused in the subsequent phase using the IAMBS algorithm presented in \cite{liu2018markov}. The IAMBS algorithm is focused on constructing an Mb for multiple variables (top $k$ nodes). It operates by determining whether the additivity property holds for Mb of $k$ variables and, further, how to proceed if the additivity property is violated by selectively performing conditional independence tests using a growing and a shrinking phase~\cite{liu2018markov}.    


\subsection{Knowledge Update Phase}
In this phase, we use the knowledge gained from the earlier phase to guide the optimization search strategy using the three components described below.

\smallskip
\noindent
\textbf{Build causal Gaussian processes.} 
At this stage, we train two surrogate models: $\text{CGP}_\text{warm}$ and $\text{CGP}_\text{cold}$ for the performance objective $\mathcal{Y}$ from $\mathcal{G}_\text{s}$ and $\mathcal{G}_\text{t}$, respectively. For this purpose, we use the mathematical formulation proposed in the CBO approach~\cite{causalBO} to build a CGP. Unlike GPs, CGPs represent the mean using interventional estimates via do-calculus, allowing the surrogate model to capture the behavior of the performance objective better than GPs (as shown in \Cref{fig:gp_vs_cgp}), particularly in areas where observational data are not available. Therefore, we fit a prior on $f(o)=E[\mathcal{Y}|do(O_i=o_i)]$ with mean and kernel function computed via do-calculus separately for each CGP obtained from $\mathcal{G}_\text{s}$ and $\mathcal{G}_\text{t}$ as the following:
\begin{equation}
    f_{e}(o) \sim GP(\mu_{e}(o),k_{c_{e}}(o, o'))
\end{equation}
\begin{equation}
    \mu_{e}(o) = \hat{E}[\mathcal{Y}|do(O_i=o_i)]
\end{equation}
\begin{equation}
    k_{c_{e}}(o, o') = k_{RBF}(o, o')+\sigma_e(o)\sigma_e(o'),
\end{equation}
where $\sigma_e(o)=\sqrt{\hat{V}_e(\mathcal{Y}|do(O_i=o_i))}$ with $\hat{V}_e$ representing the variance estimated from the configuration measurements ($\mathcal{D}_\text{s}$ or $\mathcal{D}_\text{t}$) for a particular environment. $k_{RBF}$ is the radial basis function of the kernel defined as $k_{RBF}(o,o')=exp(-\frac{||o-o'||^2}{2l^2})$, where $l$ is a hyperparameter. 
As a result, the shape of the posterior variance enables a proper calculation of the uncertainties about the causal effects (enabling identification of influential configuration options and interactions). We extract the exploration set (ES) for each environment, guided by $\mathcal{G}_\text{s}$ and $\mathcal{G}_\text{t}$, and compute the mean and uncertainty estimates for the configurations in the exploration set.

\smallskip
\noindent
\textbf{Compute acquisition function for sampling.}
Denote by $\alpha^{r}_\text{warm}(o)$ and $\alpha^{r}_\text{cold}(o)$ to be the single objective acquisition functions of the two CGPs. For \ourapproach, we choose to use expected improvement (EI) as an acquisition function~\cite{wilson2018maximizing} since EI has been shown to perform well in the configuration search. EI selects the configuration that would have the highest expected improvement to the current best interventional setting separately from $e_\text{s}$ and $e_\text{t}$ in all configurations in the respective exploration set:
\begin{equation}
    EI_{e}(o)=E_{p(y)}[max(y-y^{*},0)],
\end{equation}
where $y=E[\mathcal{Y}|do(O_i=o_i)]$ and $y^{*}$ is the optimal value observed thus far. In our implementation, we rank the configurations based on the $\alpha^{r}_\text{warm}(o)$ scores and then select the ones with the highest $\alpha^{r}_\text{cold}(o)$ score. Our acquisition function is defined as the following:
\begin{equation}
    \alpha^r(o) = \lambda^r(o)\alpha^{r}_\text{cold}(o) + (1-\lambda^r(o))\alpha^{r}_\text{warm}(o), 
\end{equation}
where $\lambda^{r}$ is an interpolation coefficient that controls the proportion of knowledge used from source and target and is dependant on $l_{\alpha}$ and the expected improvement of a configuration. The above equation shows that when $\lambda$ is 1; it would use the contribution from $\alpha_{cold}$ and use $\alpha_{warm}$ when $\lambda$ is 0. The interpolation coefficient $\lambda^{r}$ is defined as the following:
\begin{equation}
\lambda^r(o)=\mathbb{1}(\alpha^r_\text{warm*}-\alpha^r_\text{warm}(o) \leq l_{\alpha}),
\end{equation}
where $\alpha^r_\text{warm*}$ is the optimal acquisition value obtained from $\alpha^r_\text{warm}$ scores.  The choice of $l_{\alpha}$ is critical since it balances the knowledge used from the source and the target.
We set $l_{\alpha}$ to 0.1, which shows good empirical performance (as shown in \Cref{fig:rq3_a_sensitivity}. 
Intuitively, the acquisition function should operate in such a way that it uses $\alpha_{cold}$ for the configurations that are near the optimal points. Here, $l_\alpha$ is an acquisition threshold hyperparameter used to define near-optimal points w.r.t. $\alpha_{warm}$. Therefore, configurations that are closer to the optimal of $\alpha_{warm}$ (configurations that satisfy $l_{\alpha}\leq 0.1$) will provide the expected higher improvement for $\alpha_{cold}$. 

In contrast, configurations that are further away from the optimal points of $\alpha_{warm}$ (configurations that do not satisfy $l_{\alpha}\leq 0.1$) will have a higher expected improvement value for $\alpha_{warm}$. This indicates that such configurations contain options that have some environment-specific behavior that is not captured or learned correctly by the source causal model and the source causal model needs to be updated. 



\smallskip
\noindent
\textbf{Observation-intervention trade-offs.}
We find a configuration $o^\text{r+1}$ for either observation or for intervention for which the $\alpha^r$ value is maximum. We employ the $\epsilon$-greedy strategy used by the CBO to choose between observation and intervention. Observational data may be used to correctly predict the causal effects of configuration options on the performance objective. On the other hand, estimating consistent causal effects for values outside of the observable range requires intervention. The developer must identify the optimal combination of these operations to capitalize on observational data while intervening in regions with higher uncertainty. Following CBO, we define $\epsilon$ as :
\begin{equation}
    \epsilon = \frac{Vol(H(\mathcal{D}_v))}{Vol(o_{o \in \mathcal{O}}(\mathcal{D}(\mathcal{O})))} \times \frac{N}{N_{\max}},
    \label{eq:eps}
\end{equation}
where $D_v=\mathcal{D}_{\text{s}} \cup \mathcal{D}_{\text{t}}$, $Vol(H(\mathcal{D}_v))$ represents the volume of the convex hull for the observational data and $Vol(o_{o \in \mathcal{O}}(\mathcal{D}(\mathcal{O})))$ gives the volume of the interventional domain. $N_{max}$ represents the maximum number of observations the developer is willing to collect in a particular environment, and $N$ is the current size of $D_v$. The interventional space is larger than the observational space when the volume of the observational data $Vol(H(\mathcal{D}_v))$ is smaller than the number of observations $N$. Therefore, we must perform interventions to explore regions of the interventional space not covered by observational data. On the other hand, if the volume of observational data $Vol(H(\mathcal{D}_v))$ is large in relation to
$N$, we need to make observations. This is because we need to obtain consistent estimates of the causal effects, which can only be achieved with more observations. We update the convex hull incrementally for computation purposes.

\smallskip
\noindent
\textbf{Evaluate selected configuration and belief update.}
We measure the selected configuration $o^\text{r+1}$ and check whether it satisfies the constraints. If not, we replace the performance objective value with an infinitely high value to force the optimizer to avoid searching in regions of the space where the constraints are not satisfied. 
We update the causal performance and surrogate models using the new measurement. We repeat the optimization loop until the maximum budget $\beta$ is exhausted or convergence is reached, and return the configuration with minimum $\mathcal{Y}$ as optimal. 

\smallskip
\noindent

\begin{algorithm}[t]
\small
\caption{\ourapproach}
\begin{algorithmic}[1]
\REQUIRE Offline source dataset $\mathcal{D}_\text{s}$, Initial target dataset $\mathcal{D}_\text{t}$, 
Configuration space $\mathcal{O}$, Total budget $\beta$, Threshold $l_{\alpha}$, Performance Objective $\mathcal{Y}$, Constraint $\Psi$.

\rule{0.08\textwidth}{1pt} \textbf{Knowledge Extraction Phase} \rule{0.08\textwidth}{1pt}

\STATE Construct a causal performance model from $\mathcal{G}_\text{s}, \mathcal{G}_\text{t}$ using $\mathcal{D}_\text{s}, \mathcal{D}_\text{t}$, respectively.


\STATE Extract the top $k$ nodes from $\mathcal{G}_\text{s}$ in terms of causal effect on performance objective. 

\STATE Extract Markov Blanket of the top $k$ nodes to construct a new updated $\mathcal{G}_\text{s}$.

\rule{0.095\textwidth}{1pt} \textbf{Knowledge Update Phase} \rule{0.095\textwidth}{1pt}

\STATE Initialize $\text{CGP}_\text{warm}$ and $\text{CGP}_\text{cold}$.

\STATE $\beta^r = 0$

\WHILE {$\beta^{r} \leq \beta$}
\STATE Compute the exploitation coefficient $\epsilon$ using \Cref{eq:eps} and sample a random number $u \sim \mathcal{U}(0,1)$
\IF {$\epsilon > u$}
\STATE make a new observation $(o^{r+1},c^{r+1},y^{r+1})$.
\ELSE 
\STATE Set $\alpha^r_\text{warm*}=argmin_{o \in \mathcal{O}}\alpha^{r}_\text{warm}(o)$ 
\STATE Set interpolation coefficient: $\lambda^r(o)=\mathbb{1}(\alpha^r_\text{warm*}-\alpha^r_\text{warm}(o) \leq l_{\alpha})$
\STATE Set the acquisition function: $\alpha^r(o)=\lambda^r(o)\alpha^{r}_\text{cold}(o)+(1-\lambda^r(o))\alpha^{r}_\text{warm}(o)$
\STATE Pick a new configuration: $o^{r+1}=argmin_{o \in \mathcal{O}} \ \alpha^r(o)$

\STATE Intervene on the system to obtain an interventional measurement $(o^{r+1},c^{r+1},y^{r+1})$.
\ENDIF
\IF {$y^{r+1}$ does not satisfy $\Psi$}
\STATE $y^{r+1} = \infty$ 
\ENDIF
\STATE Update $\mathcal{G}_\text{s}$, $\text{CGP}_\text{warm}$, $\text{D}_\text{s}$, $\mathcal{G}_\text{t}$, $\text{CGP}_\text{cold}$, $\text{D}_\text{t}$, and $\beta^r$.

\ENDWHILE
\STATE \textbf{return} the configuration with the best performance objective.
\end{algorithmic}
 
\label{alg:cameo}
\end{algorithm}

\section{Evaluation}
\label{sec:setup}

\noindent \textbf{Subject systems and configurations.} 
We selected five configurable computer systems, including a video analytics pipeline, a \textsc{cassandra} database system, and three deep learning systems (for image, speech, and NLP, respectively). Following configuration guides and other related work~\cite{Halawa2017, iqbal2022unicorn, silva2021automatic}, we used a wide range of configuration options and system events that impact scheduling, memory management, and execution behavior. As opposed to prior works (e.g.,~\cite{velez2021white,VJSSAK:ASE20}) that only support binary options due to scalability issues, we additionally included discrete and continuous options. We use the recommended values and ranges from system documents for both of these categories of options. 

We run each software with a set of popular workloads that are extensively used in benchmarks and prototypes (more details are provided in~\Cref{sec:rq1}-\ref{sec:rq3}). We use various deployment platforms with distinct resources (e.g., computation power, memory) and microarchitectures to demonstrate our approach's versatility. We use NVIDIA Jetson \txtwo, \txone, AGX Xavier, and Xavier NX devices for edge deployment. To deploy a particular system on the cloud, we use Chameleon cloud resources where each node is a dual-socket system running Ubuntu 20.04 (GNU/Linux 6.4) with 2 Intel(R) Xeon(R) processors, 64 GB of RAM, hyperthreading, and TurboBoost. Each socket has 12 cores/24 hyper-threads with multiple Nvidia Tesla P100 16GB GPU and K80 24GB GPU for deep learning inference.  

\noindent \textbf{Data collection.} We measure the system's latency/throughput and energy for each configuration. Following a common practice~\cite{ding2021generalizable, ding2022cello}, we randomly select 2000 configurations for each system for performance measurements to determine the ground truth. We also empirically justify our selection of ground truth in Figure \ref{fig:gt_2000_vs_10000} in the appendix~\ref{sec:appendix}. We repeat each measurement $5$ times and record the median to reduce the effect of measurement noise and other variabilities~\cite{iqbal_transfer_2019}.

\noindent \textbf{Experimental parameters.}
We use a budget of 200 iterations for each optimization method, similar to standard system optimization approaches~\cite{zhang2021restune}. We repeat each method's optimization process with $3$ different random seeds for reliability. We follow the standard tuning and report parameter values for \smac, \unicorn, \restune, \restuner, and \cello. More details about experimental choices~(\Cref{tab:hw_conf}-\ref{tab:event_conf}), implementation~(\Cref{fig:hw_change_dep}-\ref{fig:severity_change_dep}), and hyperparameters~(\Cref{tab:dnn_conf}-\ref{tab:fci_conf}) are in appendix~\ref{sec:appendix}.

\noindent \textbf{Baselines.} 
We compare \ourapproach against the following:

\begin{itemize}[noitemsep, nolistsep, leftmargin=*]
    \item \textbf{\textsc{SMAC}}~\cite{hutter2011sequential}: A sequential model-based configuration optimization algorithm. 
    \item \textbf{\textsc{Unicorn}}~\cite{iqbal2022unicorn}: A method that can be used for optimization by transferring the source causal model in the target and later updating it using an active learning strategy.
    \item \textbf{\textsc{Cello}}~\cite{ding2022cello}: An optimization framework that augments Bayesian optimization with predictive early termination.
    \item \textbf{\textsc{ResTune}}~\cite{zhang2021restune}: An optimization approach that uses multiple models (ensemble) to represent prior knowledge.
    \item \textbf{\textsc{ResTune-w/o-ML}}~\cite{zhang2021restune}: \restuner without meta-learning, i.e., it only learns from scratch in the target.
\end{itemize}

\noindent \textbf{Evaluation Metrics.}
When running them for the same time limit, we compare the best performance objectives (e.g., latency, throughput, energy, etc.) achieved by each method. We also compare their relative error (RE) to present the summarized results using $  RE = \frac{|\mathcal{Y}_\text{pred}-\mathcal{Y}_\text{opt}|}{|\mathcal{Y}_\text{opt}|} \times 100\% $, where $\mathcal{Y}_\text{pred}$ is the best value achieved by each method, and $\mathcal{Y}_\text{opt}$ is the optimal measured value from our observational dataset of 2000 samples.
A method with a lower RE value is considered more effective.

\noindent \textbf{Research questions.}
We evaluate \ourapproach by answering three research questions (RQs).

\noindent\textbf{RQ1}: How effective is \ourapproach in comparison to the state-of-the-art approaches when the following environmental changes happen? (i) hardware change, (ii) workload change, (iii) software change, and (iv) deployment topology change.

\noindent\textbf{RQ2}: How does the effectiveness of \ourapproach change when the severity of environmental changes varies?

\noindent\textbf{RQ3}: How sensitive is \ourapproach when (i) the number of samples in the source environment varies? (ii) the value of $l_\alpha$ varies? and (iii) the size of the configuration space increases?



\begin{table}[tb]
\small
\caption{\small{Summarized results averaged over all environmental changes.}}
\centering
\begin{tabular}{cc|c}

\cline{1-3}
    \centering
    &\multicolumn{1}{c}{Latency}&\multicolumn{1}{c}{Energy}\\
     \cline{2-3}
    & RE(\%) &RE(\%)\\
    \cline{1-3}
   \smac &88.2&268.9\\
    \cello &46.2&182.5\\
    \restune&48.8&191.1\\
    \unicorn &55.5&179.9\\
    \restuner &29.2&81.2\\
   \hline
   \ourapproach &\cellcolor{red!20}7.8&\cellcolor{red!20}14.4\\
   \hline
\end{tabular}
\label{tab:rq1_summary}
\end{table}

  


\section{RQ1: Effectiveness in Design Explorations}
\label{sec:rq1}
We consider four types of environmental changes typically occurring when a system is deployed into production to evaluate the effectiveness of \ourapproach in finding an optimal configuration compared to the state-of-the-art.
\Cref{tab:rq1_summary} shows the summarized results for each approach averaged over different environmental changes considered in this paper. It indicates that \ourapproach outperforms other optimization approaches for both latency and energy, e.g., \ourapproach achieves 3.7$\times$ and 5.6$\times$ lower RE for latency and energy, respectively, compared to \restuner, the next best method after \ourapproach. We describe the experimental setting and the results for the four environmental changes below.

\begin{figure}[]
  \centering
  \includegraphics[width=0.48\columnwidth]{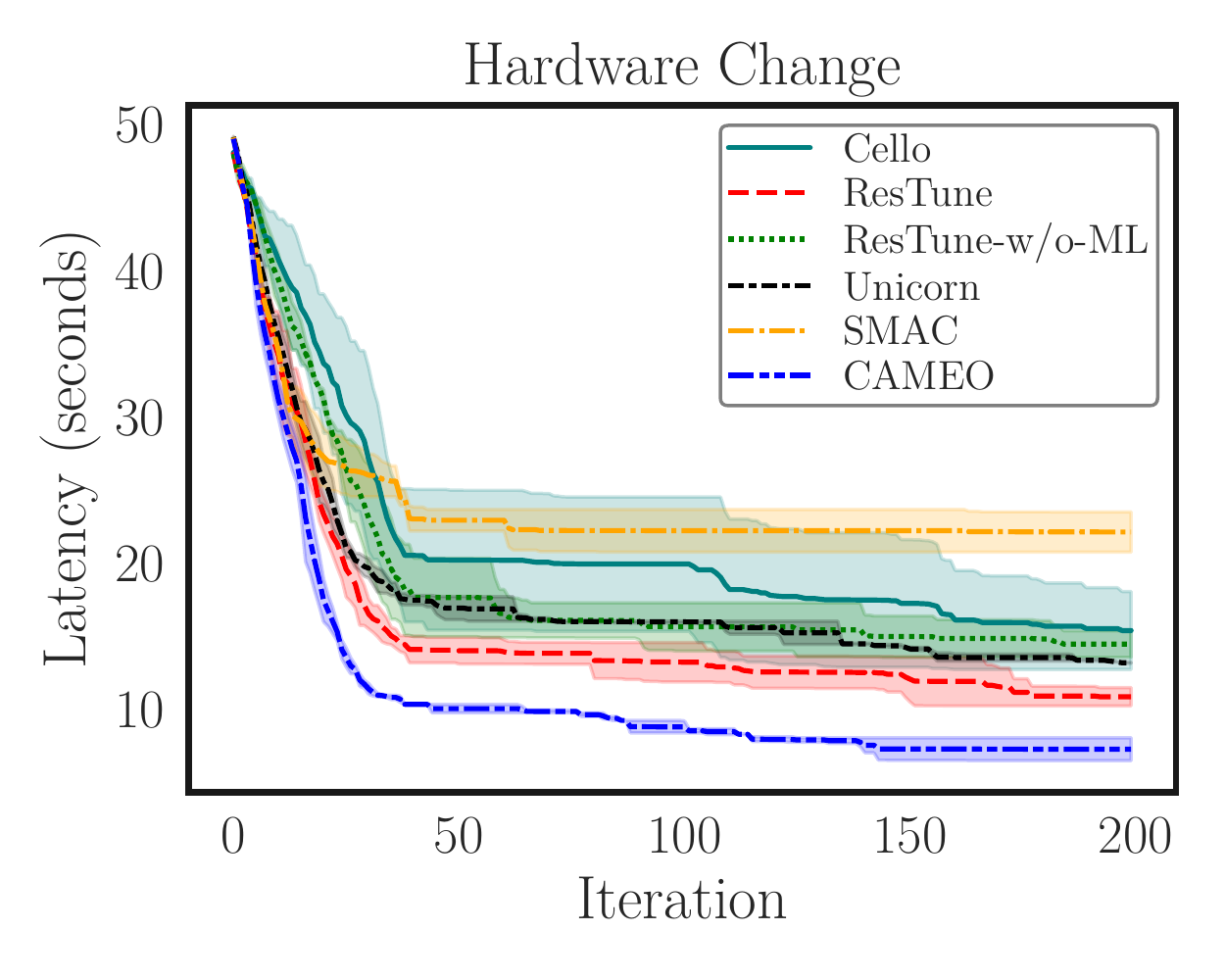}
  \includegraphics[width=0.48\columnwidth]{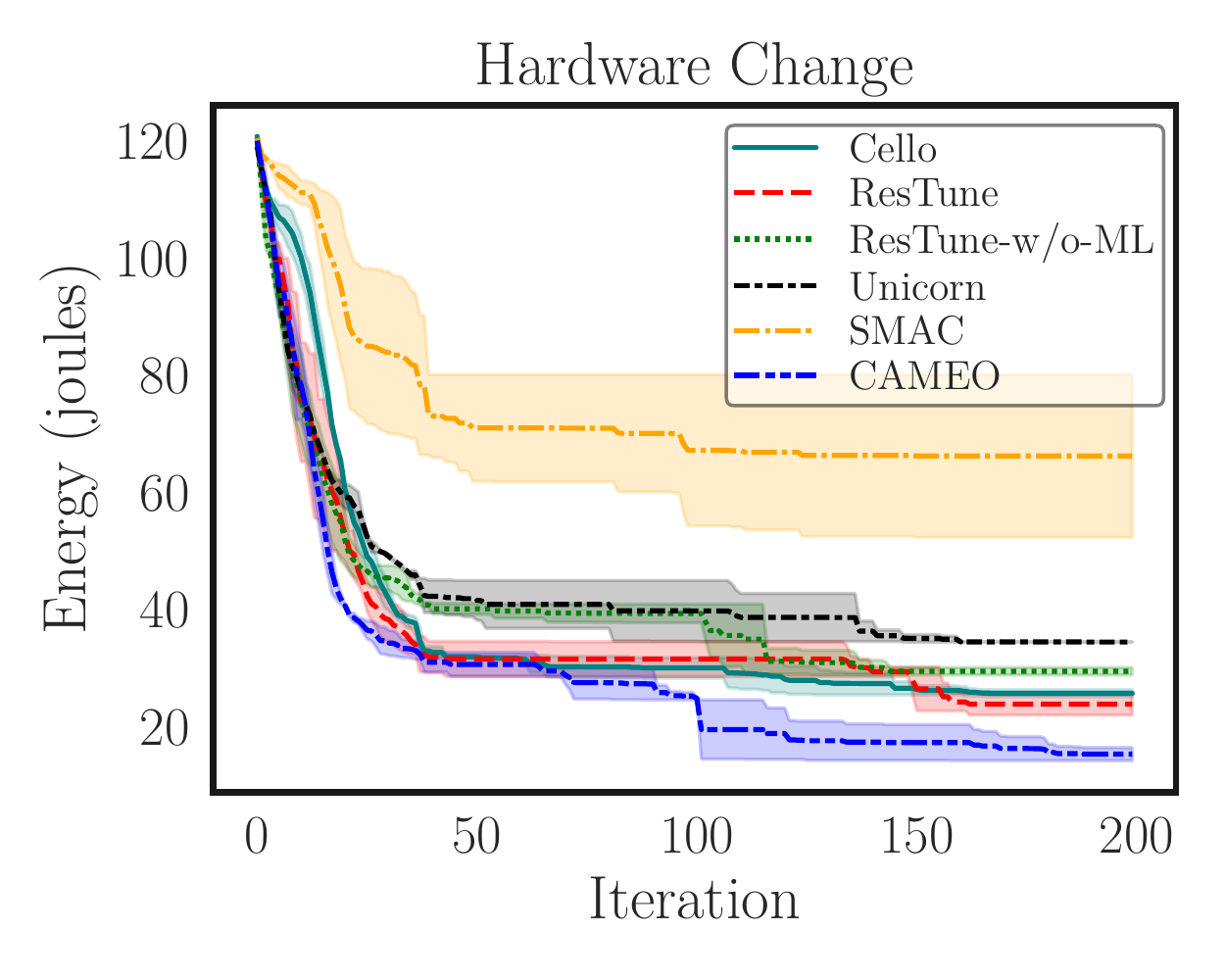}

  \caption{\small{Compared to other state-of-the-art methods, \ourapproach identifies the configurations with reduced latency (left) and energy (right) when hardware changes take place.}}
  \label{fig:rq1_hardware}
\end{figure}

\noindent \textbf{Hardware change.}
We consider the \textsc{Mlperf object detection} pipeline that uses ResNet-18 for inference of 5k images selected from the 100k test images of the ImageNet dataset~\cite{ILSVRC15}. We use \txtwo as the source hardware and \xavier as the target hardware. We examine these hardware changes since there are variable degrees of microarchitecture differences among this hardware separately. As shown in ~\Cref{fig:rq1_hardware}, \ourapproach finds the configuration with the lowest values of latency (left) and energy (right). For example, \ourapproach finds a configuration with 1.6$\times$ lower latency than \restuner. We also observe a similar trend for energy. 

\begin{figure}[t]
  \centering
  \includegraphics[width=0.48\columnwidth]{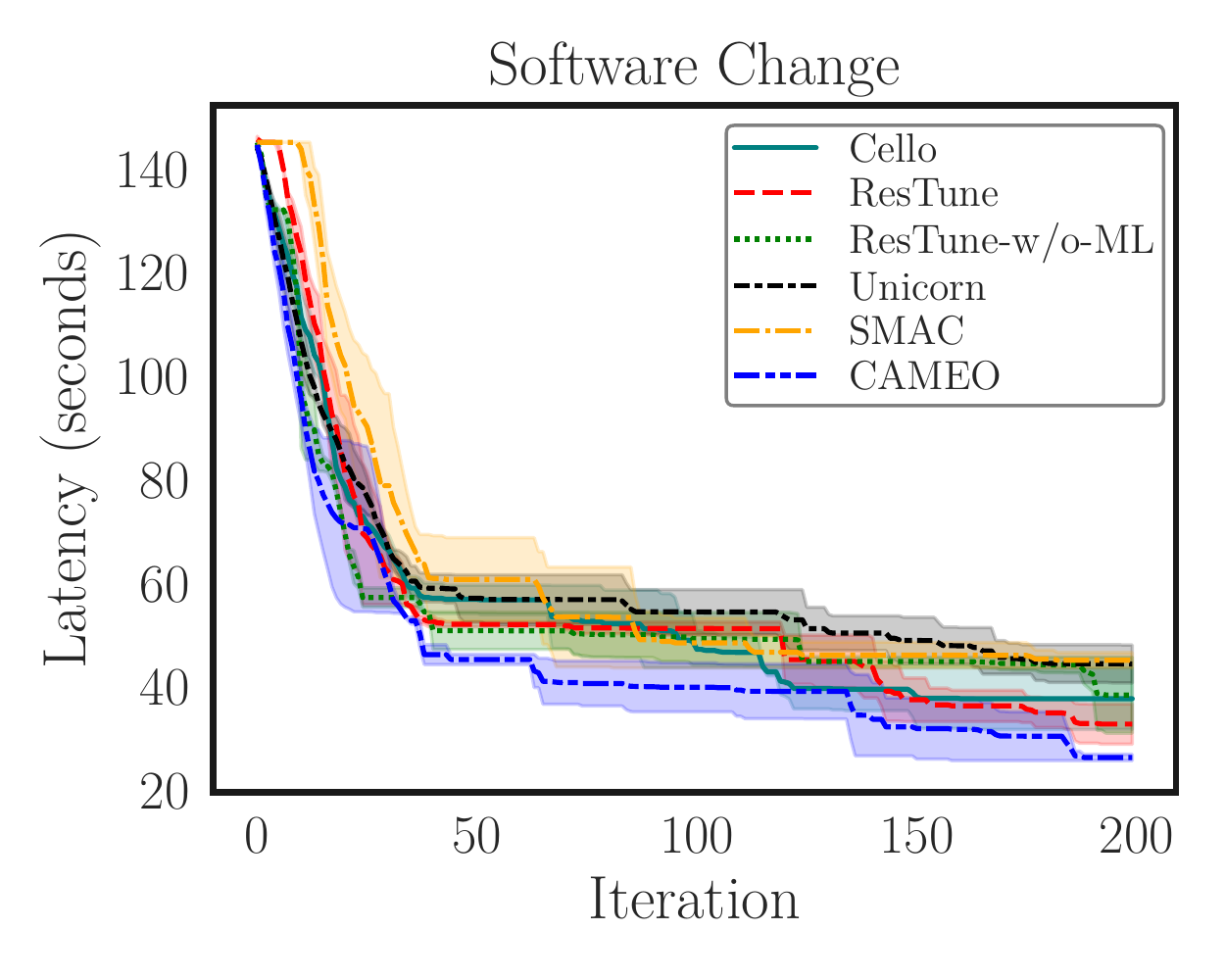}
  \includegraphics[width=0.48\columnwidth]{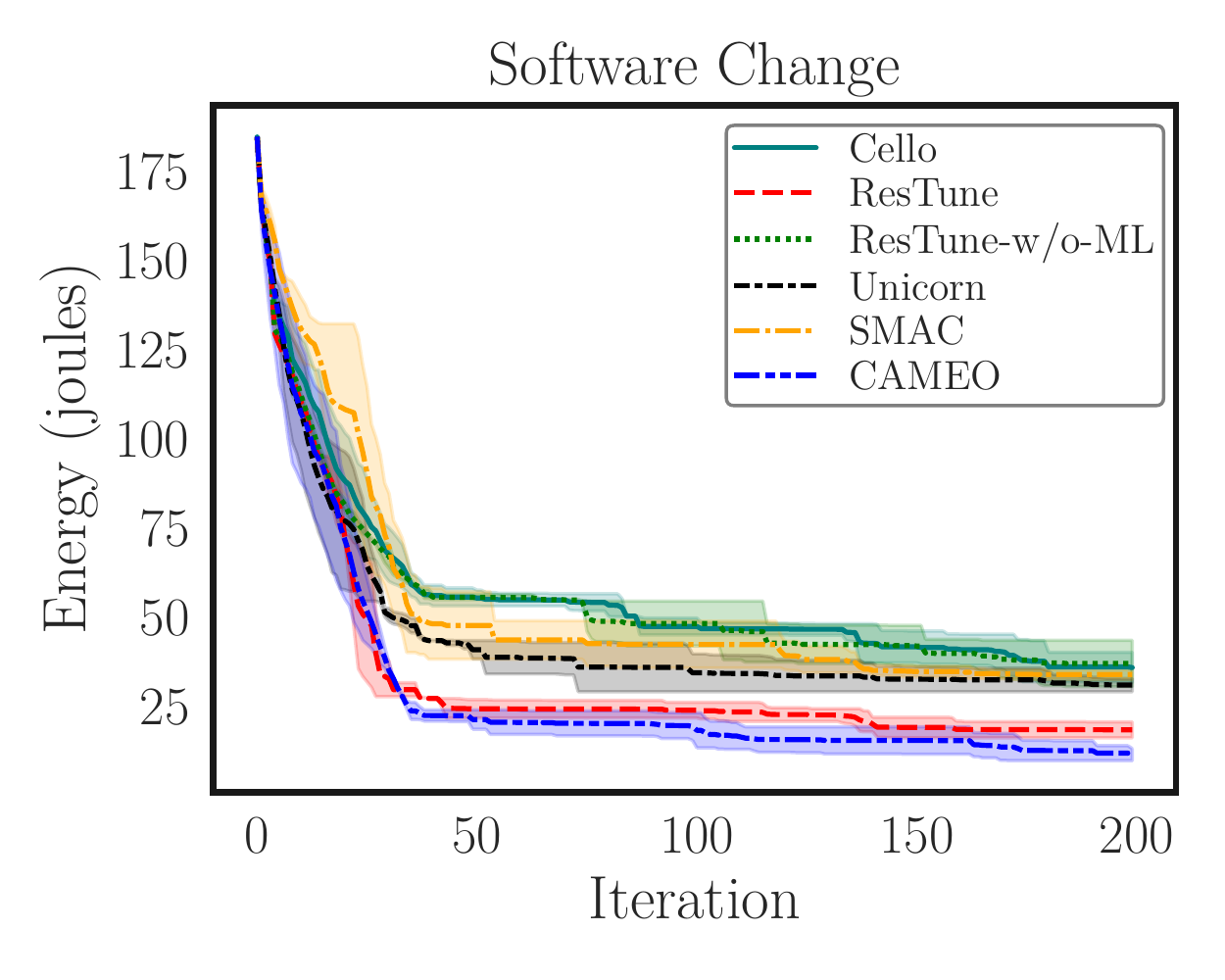}
  \caption{\small{\ourapproach finds the configurations with lowest latency (left) and energy (right) when software changes.}}
   \label{fig:rq1_software}
\end{figure}

\noindent \textbf{Software change.}
We consider variants of a natural language processing (NLP) model---\textsc{BERT}~\cite{devlin2018bert} and \textsc{TinyBERT}~\cite{jiao2019tinybert}---deployed on \xavier in our experiments. We set up a software change by changing the model architecture across environments, where we use \textsc{TinyBERT} with 3 million parameters as the source and \textsc{BERT-Base} with 109M parameters as the target. As a workload, we perform sentiment analysis on 1000 of the 25,000 reviews from the IMDB test dataset~\cite{maas-EtAl:2011:ACL-HLT2011}. The results presented in \Cref{fig:rq1_software} demonstrate that the optimal configurations found by \ourapproach have a 1.1$\times$ lower latency and a 1.7$\times$ lower energy value compared to \restuner.

\begin{figure}[t]
  \centering
  \includegraphics[width=0.48\columnwidth]{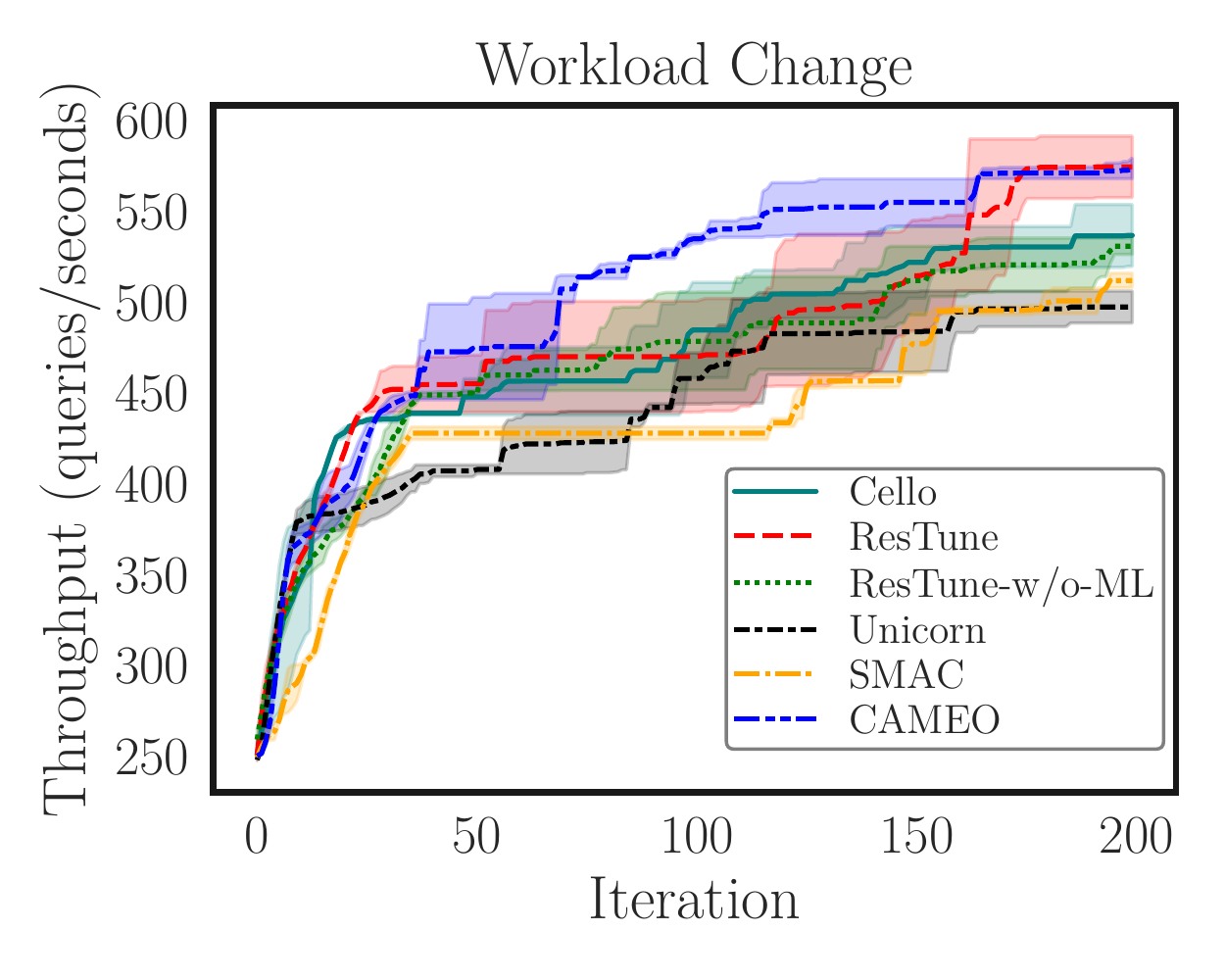}
  \includegraphics[width=0.48\columnwidth]{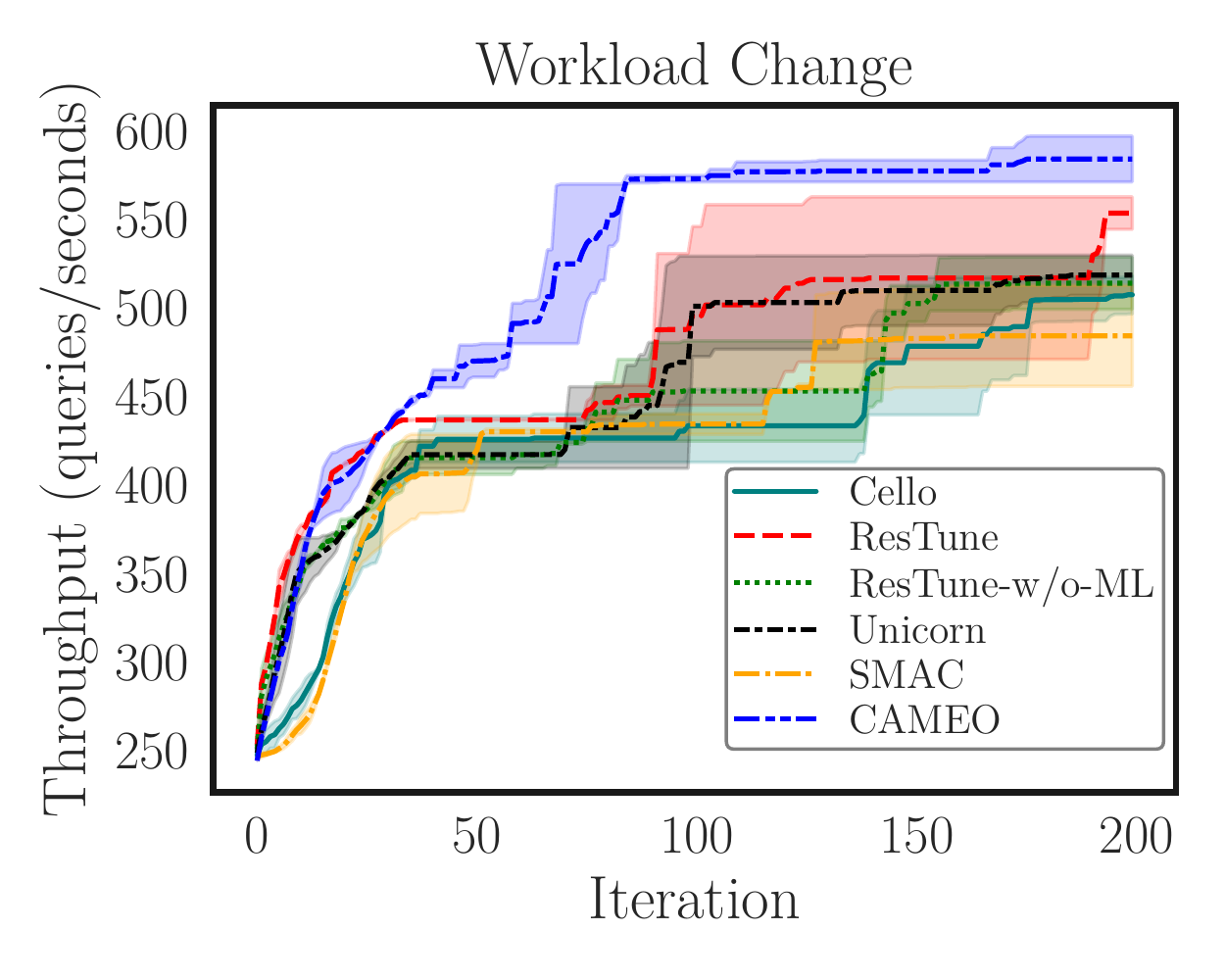}

  \caption{\small{For read-only to balanced workload change, \restuner slightly outperforms \ourapproach in finding configurations with higher throughput (left). For read-only to update-heavy workload change, \ourapproach dominates other approaches in finding optimal configuration higher throughput (right).}}
   \label{fig:rq1_workload}
\end{figure}

\noindent \textbf{Workload change.}
We consider \textsc{Cassandra} database deployed on \chameleon (see Section ~\ref{sec:setup}) while varying different workloads to create different source and target environments using the TPC-C benchmark~\cite{tpc-benchmark}. We use a YCSB workload generator to generate 3 workloads: (i) \textsc{read only}- 100\% read, (ii) \textsc{balanced} - 50\% read and 50\% update, and (iii) \textsc{update heavy} - 95\% update and 5\% read. To optimize throughput, we use a \textsc{read only} workload as the source and the remaining two workloads as the target separately. Results for workload changes are presented in \Cref{fig:rq1_workload}. When the workload changes from \textsc{read only} to \textsc{balanced},  \restuner outperforms \ourapproach by finding a configuration with 1.02$\times$  higher throughput. Upon further investigation, we found that the distributions between source and target were relatively similar, and the shared covariance learning in \restuner helped to find a better configuration. Additionally, the knowledge extraction module in \restuner is particularly developed to correctly capture workload behavior, making it more suitable for this workload change scenario. However, as the distribution difference increases, \ourapproach outperforms \restuner, e.g., for \textsc{update heavy} workload \ourapproach has 1.06$\times$ higher throughput than \restuner.

\begin{figure}[t]
  \centering
  \includegraphics[width=0.48\columnwidth]{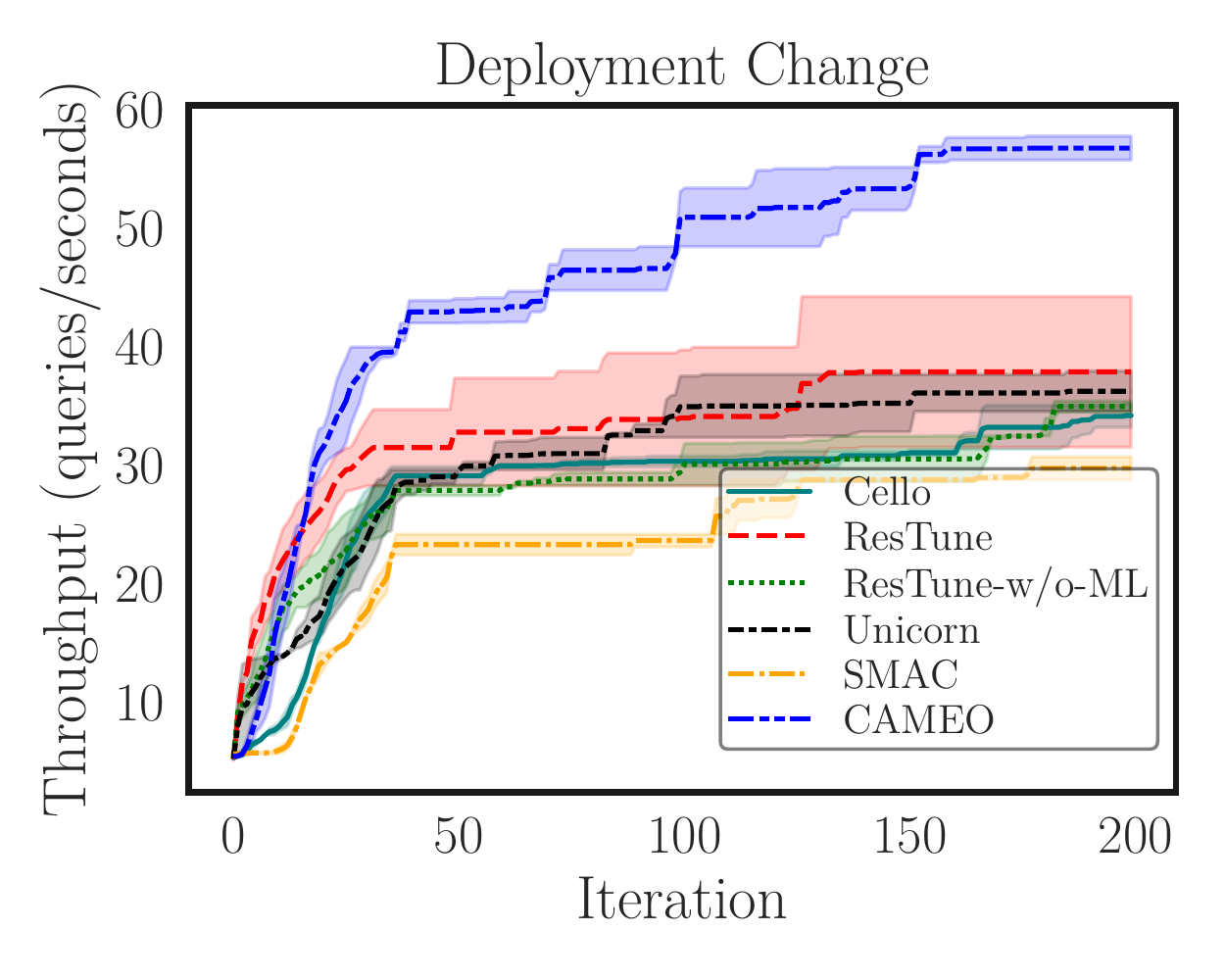}
  \includegraphics[width=0.48\columnwidth]{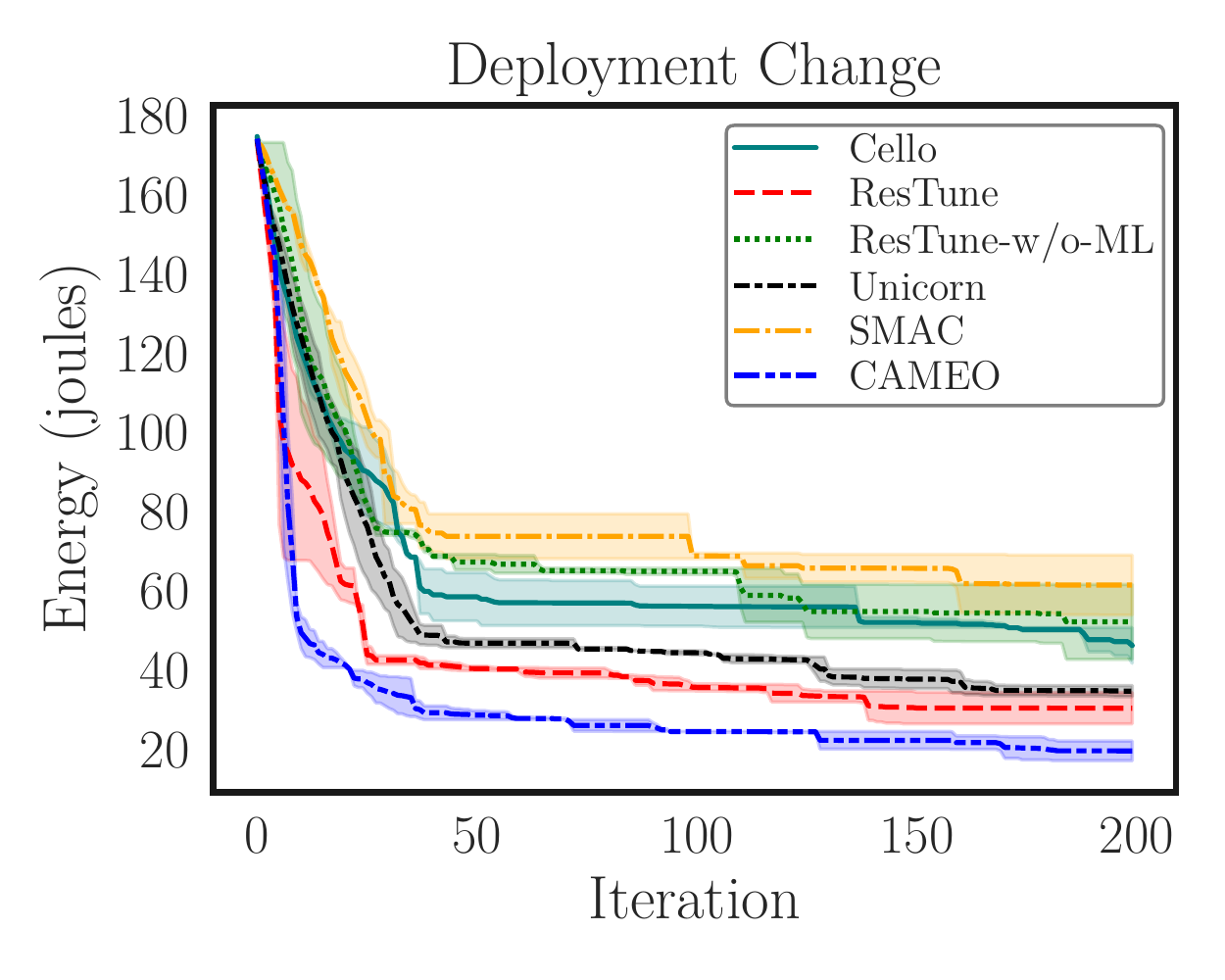}

  \caption{\small{\ourapproach has maximum effectiveness in finding configurations with the highest throughput (left) and energy (right) when deployment topology changes.}}
   \label{fig:rq1_deployment}
\end{figure}

\noindent \textbf{Deployment topology change.}
To test the effectiveness of \ourapproach across deployment topology change, we consider a video analytics pipeline: \deepstream that uses 4 camera streams as the workload. Our \deepstream pipeline has four components: (i) an x264 decoder, (ii) a multiplexer, (iii) a TrafficCamNet model with ResNet-18 as the detector, and (iv) an NvDCF tracker, which uses a correlation filter-based discriminative learning algorithm online for tracking. As the source environment, we adopt a centralized deployment topology in which all four components run on \xavier NX hardware. For the target, we use a distributed deployment topology with two \xavier NX hardware, deploying the decoder and multiplexer in one and the detector and tracker in the other. We use \textsc{Apache Kafka} to send and receive the output of the multiplexer to the detector that uses a binary protocol over TCP. Our experimental results for the changes in the deployment environment (Figure~\ref{fig:rq1_deployment}) show that \ourapproach significantly outperforms others in finding the optimal throughput and energy. For example, the optimal configuration discovered by \ourapproach has an improvement of 1.3$\times$ and 1.5$\times$ (as) for throughput and energy, respectively, than the next-best method. 

\noindent \textbf{Summary of observations.}
Methods based on guided knowledge reuse (\ourapproach and \restuner) consistently are the top performers over methods that do not reuse knowledge. The steep performance curves during the earlier iterations indicate that the optimization process's warm-starting helps quickly go to the region containing good configurations. As a result, in all environmental changes, methods that reuse knowledge from the source outperform \smac, \restune, and \cello that do not rely on previous information and cannot achieve the optimal within the allowed budget. Unlike \restuner and \ourapproach, \unicorn directly uses source information in the target, thereby introducing bias, which must be learned. This unlearning is not necessary for \ourapproach due to its knowledge transfer strategy.

\begin{table}[b]
\scriptsize
\caption{\small{Optimal configuration discovered by different baselines. Configuration options are ranked in descending order based on their average causal effect (ACE) value on the performance objective, i.e., Latency.}}
\centering
\resizebox{\columnwidth}{!}{
\begin{tabular}{cccccccccc}
\hline
Configuration& \smac & \unicorn & \restuner &\restuner&\ourapproach&ACE&Optimal\\
Option&&&\textsc{-w/o-ML}&&&&\\
\hline
\textsf{cpu\_frequency}&1.3&1.6&1.6&1.6&\cellcolor{red!20}2.0&0.19&\cellcolor{red!20}2.0\\
\textsf{vm.dirty\_ratio}&20&\cellcolor{red!20}5&20&\cellcolor{red!20}5&\cellcolor{red!20}5&0.13&\cellcolor{red!20}5\\
\textsf{vm.swappiness}&\cellcolor{red!20}60&\cellcolor{red!20}60&\cellcolor{red!20}60&\cellcolor{red!20}60&\cellcolor{red!20}60&0.11&\cellcolor{red!20}60\\
\textsf{gpu\_frequency}&\cellcolor{red!20}1.3&\cellcolor{red!20}1.3&\cellcolor{red!20}1.3&\cellcolor{red!20}1.3&\cellcolor{red!20}1.3&0.08&\cellcolor{red!20}1.3\\
\textsf{num\_cores}&3&\cellcolor{red!20}4&3&\cellcolor{red!20}4&\cellcolor{red!20}4&0.06&\cellcolor{red!20}4\\
\textsf{memory\_growth}&0.5&0.9&0.5&0.9&\cellcolor{red!20}-1&0.04&\cellcolor{red!20}-1\\
\textsf{emc\_frequency}&1.1&\cellcolor{red!20}1.3&\cellcolor{red!20}1.3&1.1&\cellcolor{red!20}1.3&0.009&\cellcolor{red!20}1.3\\
\textsf{drop\_caches}&\cellcolor{red!20}0&\cellcolor{red!20}0&\cellcolor{red!20}0&\cellcolor{red!20}0&\cellcolor{red!20}0&0.008&\cellcolor{red!20}0\\
\textsf{scheduler\_policy}&\cellcolor{red!20}NOOP&\cellcolor{red!20}NOOP&CFP&\cellcolor{red!20}NOOP&\cellcolor{red!20}NOOP&0.001&\cellcolor{red!20}NOOP\\
\textsf{vm.vfs\_cache\_pressure}&\cellcolor{red!20}10&50&\cellcolor{red!20}10&\cellcolor{red!20}10&\cellcolor{red!20}10&0.001&\cellcolor{red!20}10\\
\textsf{vm.dirty\_bytes}&\cellcolor{red!20}30&60&60&\cellcolor{red!20}30&60&0.0009&\cellcolor{red!20}30\\
\textsf{kernel.sched\_rt\_runtime\_us}&5x10$^6$&5x10$^6$&5x10$^6$&5x10$^6$&\cellcolor{red!20}9.5x10$^6$&0.0009&\cellcolor{red!20}95x10$^6$\\
\textsf{logical\_devices}&\cellcolor{red!20}1&\cellcolor{red!20}1&\cellcolor{red!20}0&\cellcolor{red!20}1&\cellcolor{red!20}1&0.0008&\cellcolor{red!20}1\\
\textsf{kernel.sched\_child\_runs\_first}&\cellcolor{red!20}0&\cellcolor{red!20}0&\cellcolor{red!20}0&\cellcolor{red!20}0&\cellcolor{red!20}0&0.0006&\cellcolor{red!20}0\\
Latency&22s&15s&14s&13s&8s&&8s\\

\hline
\end{tabular}}
\label{tab:optimal_config}
\end{table}





\noindent \textbf{Why \ourapproach works better?} 
To further explain \ourapproach's advantages over other methods, we
conduct a case study using the same experimental setup mentioned in \Cref{sec:motivation} where \textsc{Mlperf Object Detection} pipeline is deployed on \txtwo as the source and \xavier as the target. We discuss our key findings in the following.

\begin{figure}[ht]
\centering
\includegraphics[width=0.48\columnwidth]{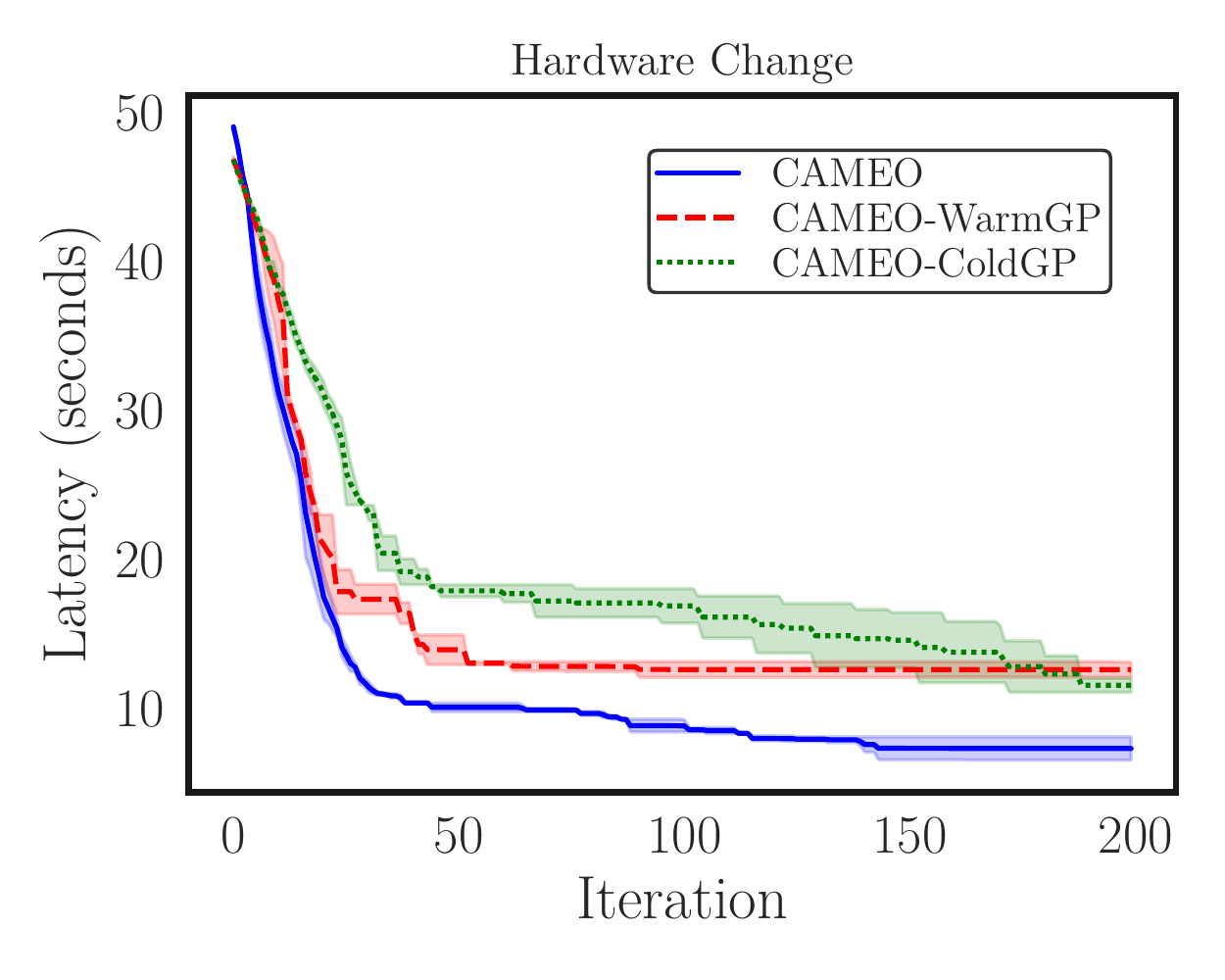}
\includegraphics[width=0.48\columnwidth]{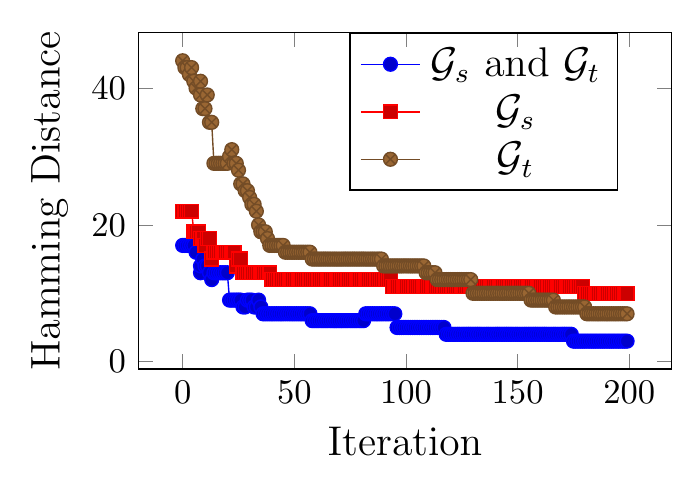}
\caption{\small{The causal performance models become more accurate with increasing iterations. The correctness of $\mathcal{G}_\text{s}$ and $\mathcal{G}_\text{t}$ when combined helps \ourapproach in detecting the optimal configuration more effectively than others. A lower hamming distance value indicates a smaller difference with the ground truth causal performance model in the target.}}
\label{fig:shd_cgp}

\end{figure}

\noindent \textbf{(i) The combined correctness of two causal performance models allows one to effectively identify the values of optimal options.} Table~\ref{tab:optimal_config} shows the optimal configuration discovered by different approaches. It is evident that \ourapproach can correctly identify the maximum number of options values compared to other approaches (only misidentified \textsf{vm.dirty\_bytes}). This is possible due to the usage of two causal models $\mathcal{G}_\text{s}$ and $\mathcal{G}_\text{t}$ as shown on the left of Figure~\ref{fig:shd_cgp}. The right subfigure in Figure~\ref{fig:shd_cgp} shows the iterative changes in structural differences (by Hamming distance) with the causal model of the ground truth when using only $\mathcal{G}_\text{s}$ or $\mathcal{G}_\text{t}$ or when combining the two. Here, we find that the Hamming distance is significantly low when both $\mathcal{G}_\text{s}$ and $\mathcal{G}_\text{t}$ are combined, indicating that the discovered causal performance model is nearly identical to the ground truth causal performance model in the target, as shown in Figure~\ref{fig:shd_cgp}.

\begin{figure}[ht]
\centering

\includegraphics[width=\linewidth]{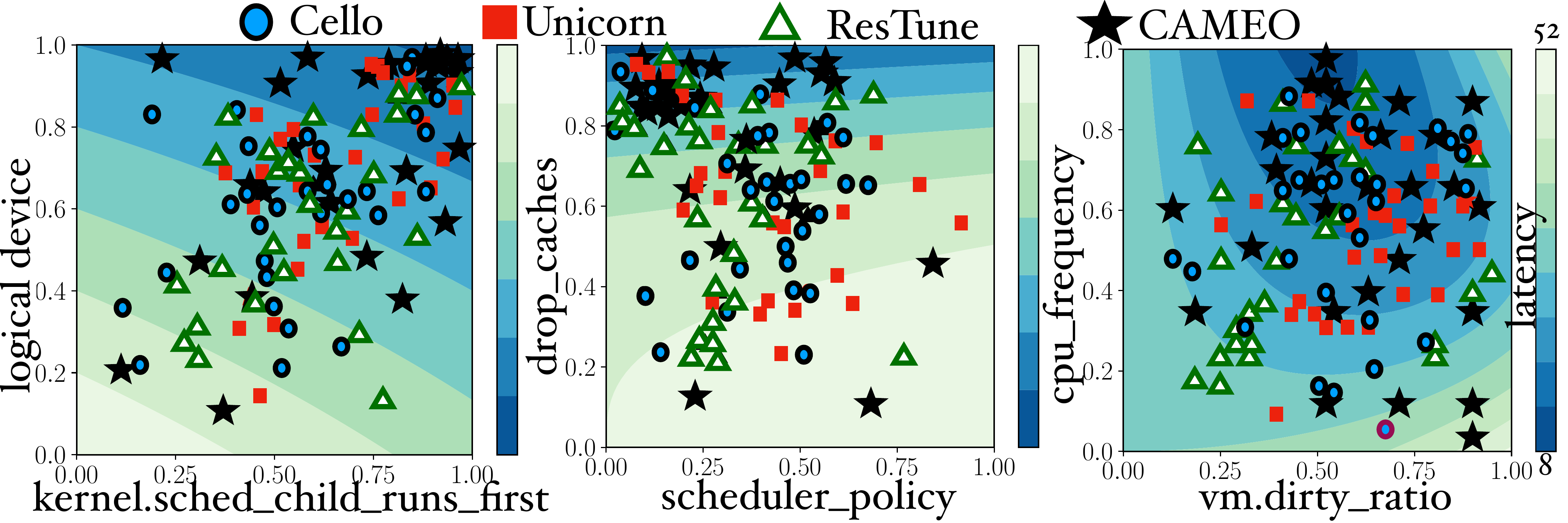}
\caption{\small{Contour plot with options of different causal effects. The color bar indicates the latency values, where lower values indicate better performance.}}
\label{fig:cameo_better}
\end{figure}


\noindent \textbf{(ii) \ourapproach has utilized the budget more efficiently by carefully evaluating core configuration options.} 
To better understand the optimization process, we visualize the response surfaces of three sets of options pairwise with different degrees of average causal effect (ACE) on latency (Figure~\ref{fig:cameo_better}). The leftmost subfigure of Figure~\ref{fig:cameo_better} contains options with lower ACE values, while the rightmost contains the options with high ACE values only). The middle subfigure of Figure~\ref{fig:cameo_better} contains options that have ACE values near the median (the ACE values of the configuration options are provided in Table~\ref{tab:optimal_config}). The right-hand subfigure of Figure~\ref{fig:cameo_better} shows that the response surfaces of the options with higher ACE values are more complex than those with lower ACE values. Table \ref{tab:optimal_config} demonstrates that \ourapproach can accurately find the optimal values of options with higher ACE values, such as \textsf{cpu\_frequency} and \textsf{dirty\_ratio}, demonstrating a better understanding of such complex behavior. Figure~\ref{fig:cameo_better} also shows how \ourapproach has investigated more configurations by varying more options with higher ACE values than lower ones. By focusing on more sophisticated surfaces rather than wasting resources on less effective options, \ourapproach can make the best use of resources to better understand the performance behavior for navigating the search space.

\noindent \textbf{(iii) \ourapproach reaches better configurations by achieving better exploration-exploitation trade-offs.} From Figure~\ref{fig:cameo_better} (left and middle), we observe that for options with lower ACE values, \ourapproach quickly reaches the region with configurations with lower latency within fewer explorations and then focuses on exploitation behavior to quickly determine the optimal configuration.  In the rightmost subfigure of Figure~\ref{fig:cameo_better}, configurations evaluated by \ourapproach cover the largest number of different regions (indicating a better exploration). Here, we also observe that \ourapproach has evaluated a higher number of configurations near the optimal configuration (blue) regions of the response surface (indicating better exploitation). Therefore, \ourapproach has a higher coverage of the configurations evaluated during the optimization procedure compared to other approaches for the core options with higher ACE values. The identification of such core features is central to achieving better exploration-exploitation trade-offs.

\section{RQ2: Severity of Environmental Changes}
\label{sec:rq2}

\begin{figure*}[t]
  \centering
  \includegraphics[width=0.32\linewidth]{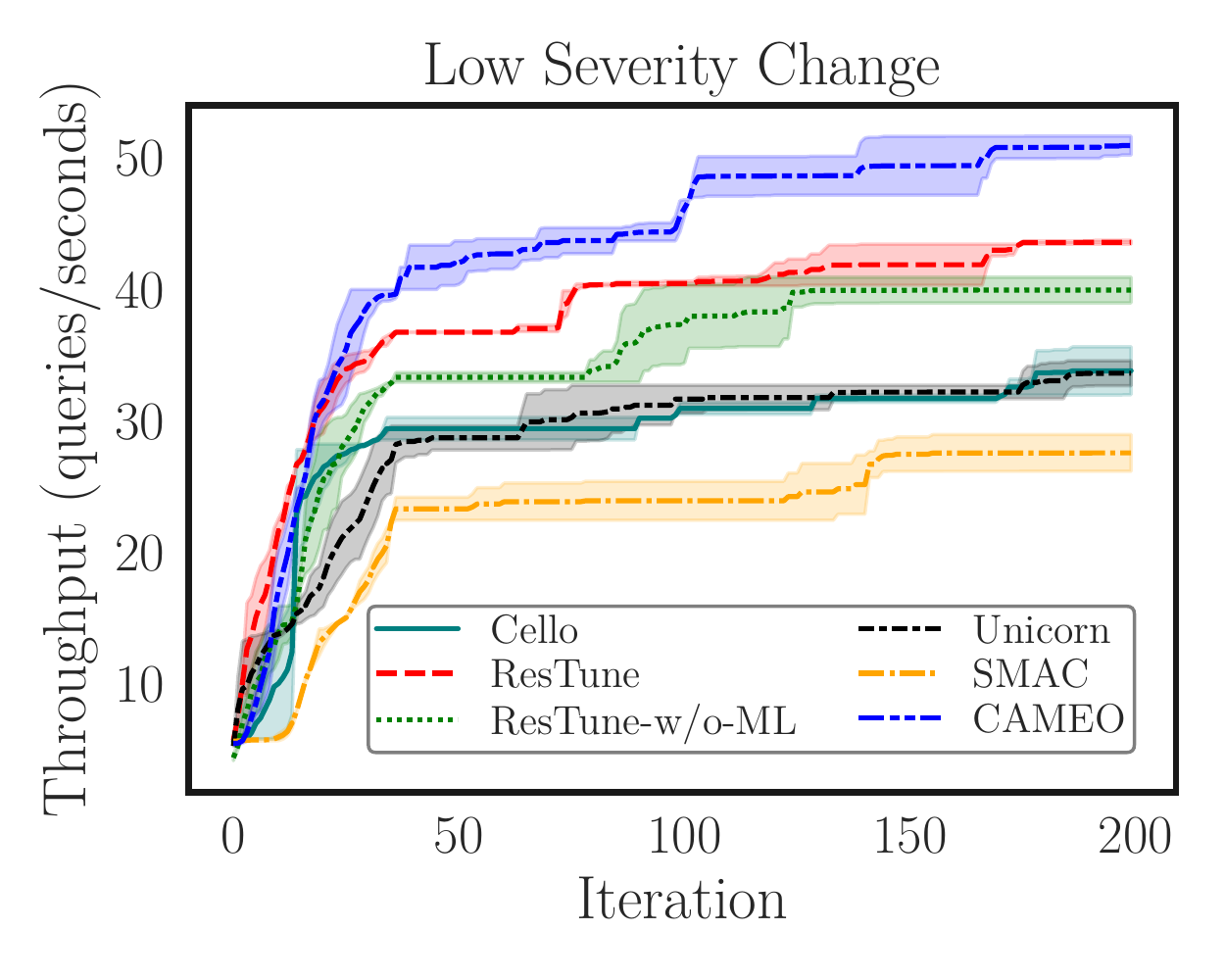}
  \includegraphics[width=0.32\linewidth]{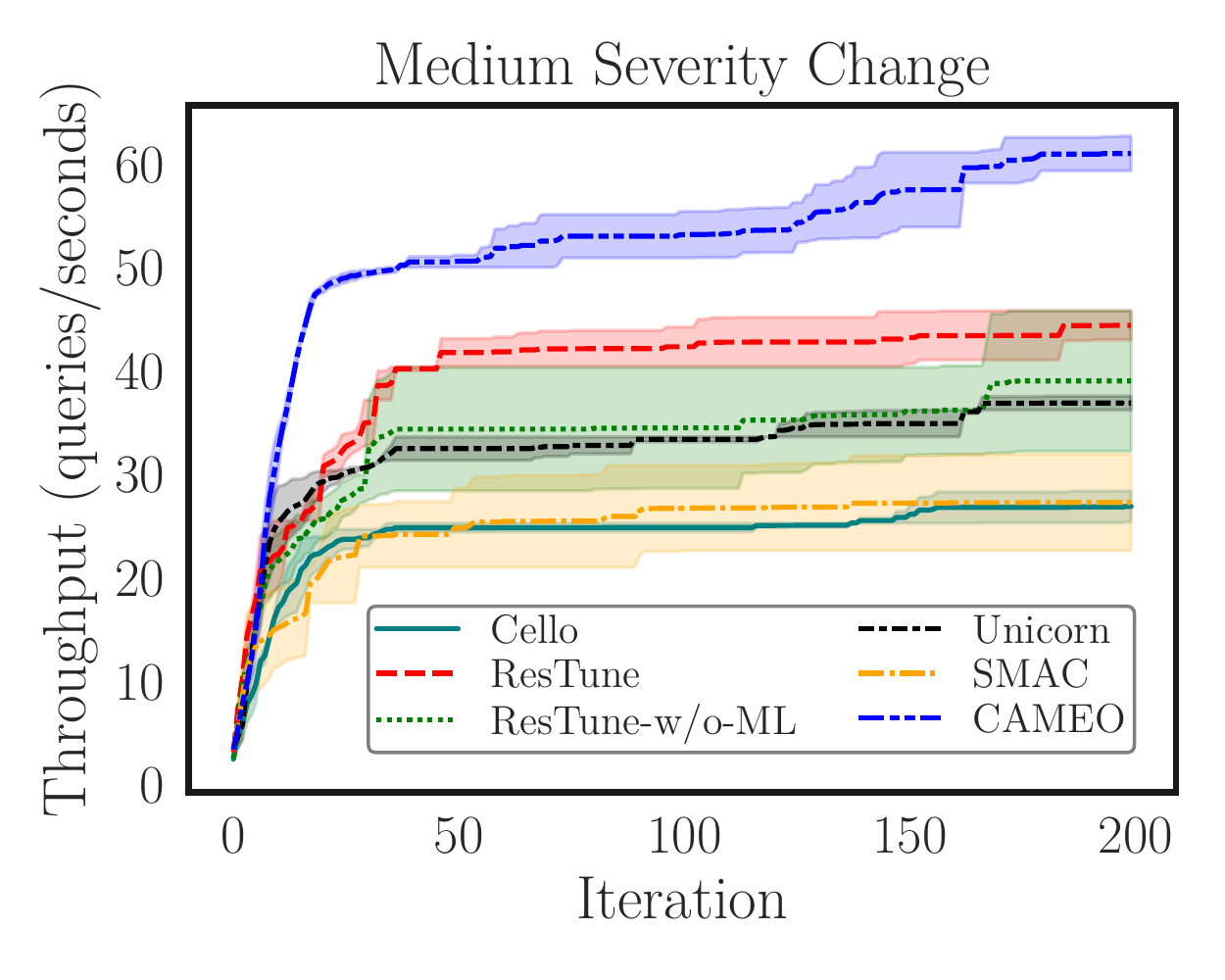}
  \includegraphics[width=0.32\linewidth]{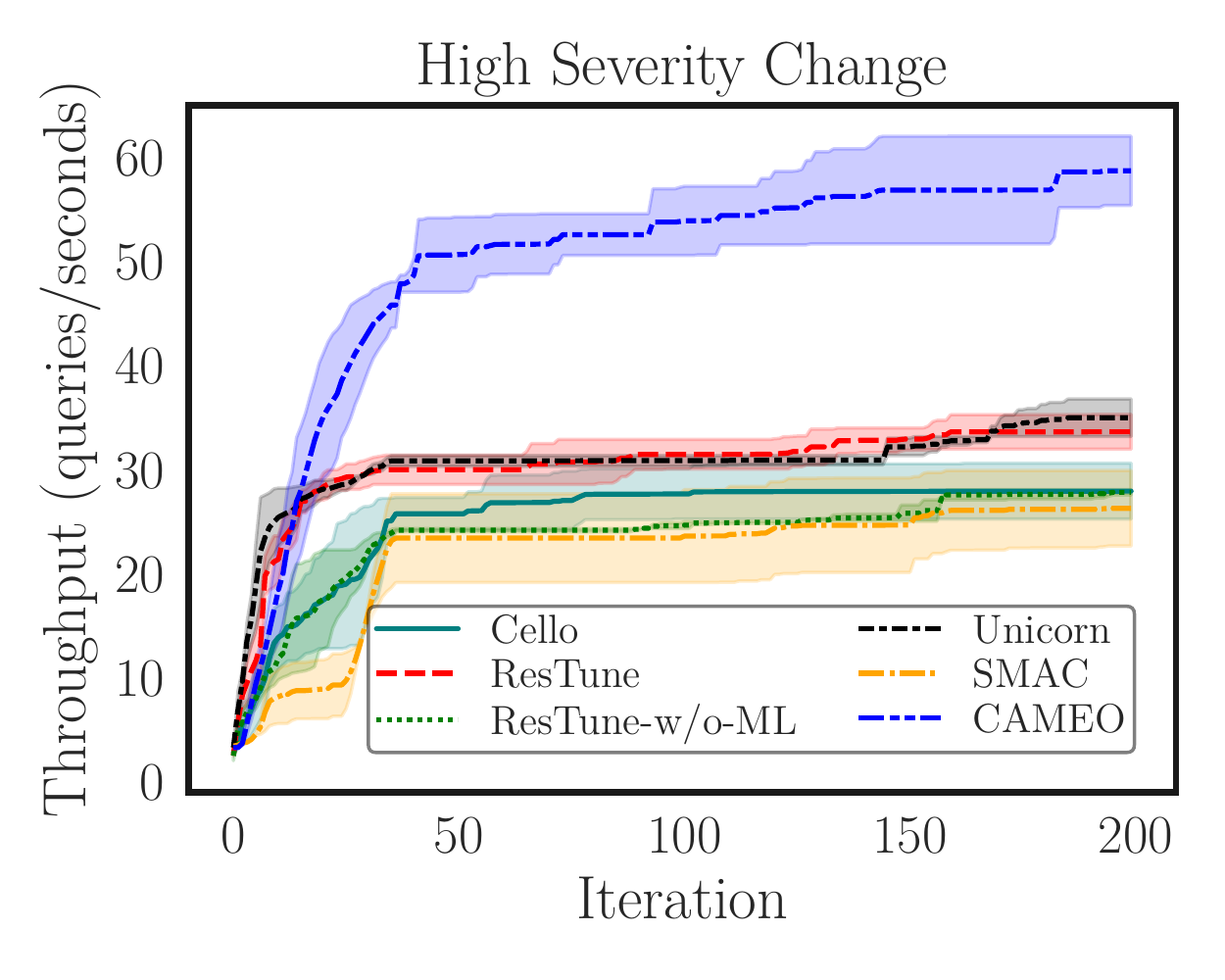}
  \caption{\small{\ourapproach achieves higher throughput when different severity of environmental changes take place.}}
  \label{fig:rq2_severity}
\end{figure*}
The effectiveness of \ourapproach changes due to the amount of distribution shift during environmental changes. Predicting how much the distribution will change when an environmental change occurs is impossible. Therefore, it is critical to understand how sensitive \ourapproach is to different degrees of severity of change. Following previous work~\cite{JSVKPA:ASE17}, we consider various environmental changes of varying severity to answer this question. The scale and the number of changes that occur indicate the severity. For example, an environment change is more severe if both hardware and workload change, compared with only hardware changes.

We consider the centralized deployment of \deepstream used in RQ1 as the source and use the following as the targets: 
(i) \textit{Low severity}: We only change one category, hardware (AGX \xavier to \xavier NX);
(ii) \textit{Medium severity}: We consider the change of two categories, hardware and deployment topology. In this setup, the target is deployed with \deepstream in a distributed fashion on two \xavier NX devices with a decoder with four camera streams as workload; and
(iii) \textit{High severity}: We consider a change of four categories, workload, deployment topology, hardware, and model. Our target has \deepstream distributedly deployed on two TX2s, with a workload of eight camera streams. We also changed the detector from ResNet-18 to ResNet-50. 


\noindent \textbf{Results.}
As shown in \Cref{fig:rq2_severity}, \ourapproach constantly outperforms the baselines by achieving maximum throughput for all severity of environmental changes. For example, \ourapproach finds a configuration with 1.3$\times$, 1.5$\times$, and 1.9$\times$ higher throughput than \restuner with low, medium, and high severity of changes, respectively. The KL divergence values between the distributions of the source and the low, medium, and high severity environmental changes setup are 418, 951, and 1329. Therefore, we conclude that \ourapproach performs better than the baselines as environmental changes become more severe.

\section{RQ3: Sensitivity and Scalability}
\label{sec:rq3}
First, we investigate \ourapproach's performance under different source measurements and how this affects the knowledge transferred from the source to the target and overall performance. Second, we determine how the value of $l_\alpha$ influences \ourapproach's effectiveness. Finally, we investigate \ourapproach's scalability in larger 
configuration space. 
\begin{figure}[t]
  \centering
\includegraphics[width=0.48\linewidth]{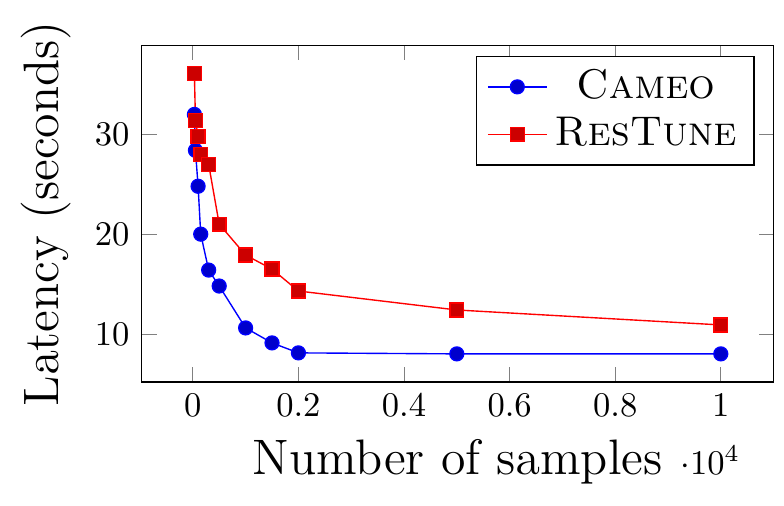}
{ \includegraphics[width=0.48\linewidth]{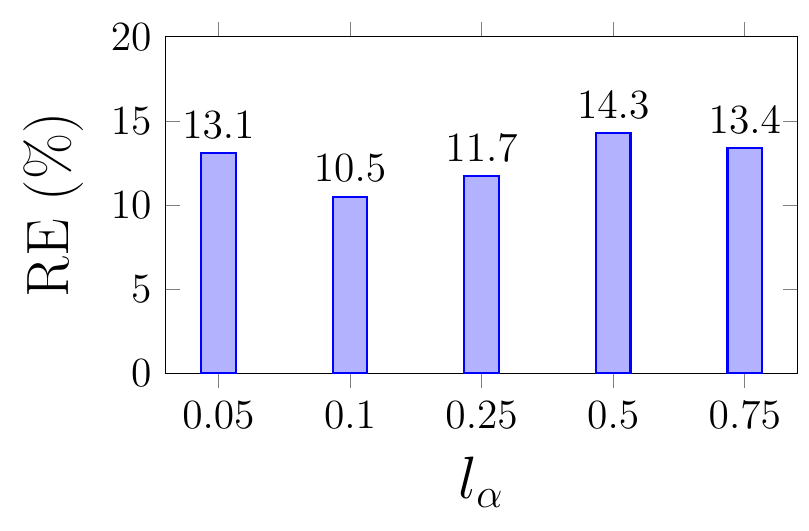}}
  \caption{\small{ (left) Both approaches find better configurations with higher source samples. Compared to \restuner, \ourapproach finds the optimal configuration with a lower minimum latency (right). \ourapproach has a minimum RE when $l_\alpha$ is 0.1. }}
   \label{fig:rq3_a_sensitivity}
\end{figure}

\noindent \textbf{Sensitivity to the number of source measurements.}
We consider the \modp pipeline deployed in \txtwo as the source and the same pipeline in \xavier as the target, varying the number of measurements in \txtwo from 30 to 10000 for evaluation and comparison of their optimal values discovered by different approaches. As shown in \Cref{fig:rq3_a_sensitivity} (left), increasing the number of source measurements positively influences \ourapproach's as compared to \restuner. Including a greater number of source samples increases the danger of bias from the source environment, particularly when the distributions of two environments are extremely disparate. From this figure, we can infer that \ourapproach can prevent those biases from being introduced into the target because more samples are used to extract knowledge from the source. We also observe that \ourapproach reaches a plateau (after 2000 samples) faster than \restuner, indicating that \ourapproach can find better configurations with fewer source samples. Because of \ourapproach's ability to detect the core features, it can be reliably used across environments without much modification.

\begin{table}[b]
\scriptsize
\centering
\caption{\small{Comparison of computation time in seconds per iteration for baselines compared to \ourapproach. Lower is better.}}
\resizebox{.45\textwidth}{!}{
\begin{tabular}{cccc}
\hline
 & Model & Configuration& Total \\
Method & Update &Recommendation & Time \\
& Time & Time &\\
\hline 
\smac  & \cellcolor{red!20}5.6 & \cellcolor{red!20}9.2 & \cellcolor{red!20}58.1           \\
\cello  & 8.1 & \cellcolor{red!20}9.2 & 60.3          \\ 
\unicorn  & 11.5 & 11.3 & 65.4          \\
\restune  & 8.3 & \cellcolor{red!20}9.2 & 61.3          \\
\restuner  & 9.7 & 9.7 & 63.4          \\
\hline
\ourapproach & 12.7 & 14.4 & 71.6          \\

\hline

\end{tabular}}
\label{tab:compute_time}
\end{table}

\noindent \textbf{Scalability to the number of configuration options.}
\begin{figure}
  \centering
{ \includegraphics[width=0.48\columnwidth]{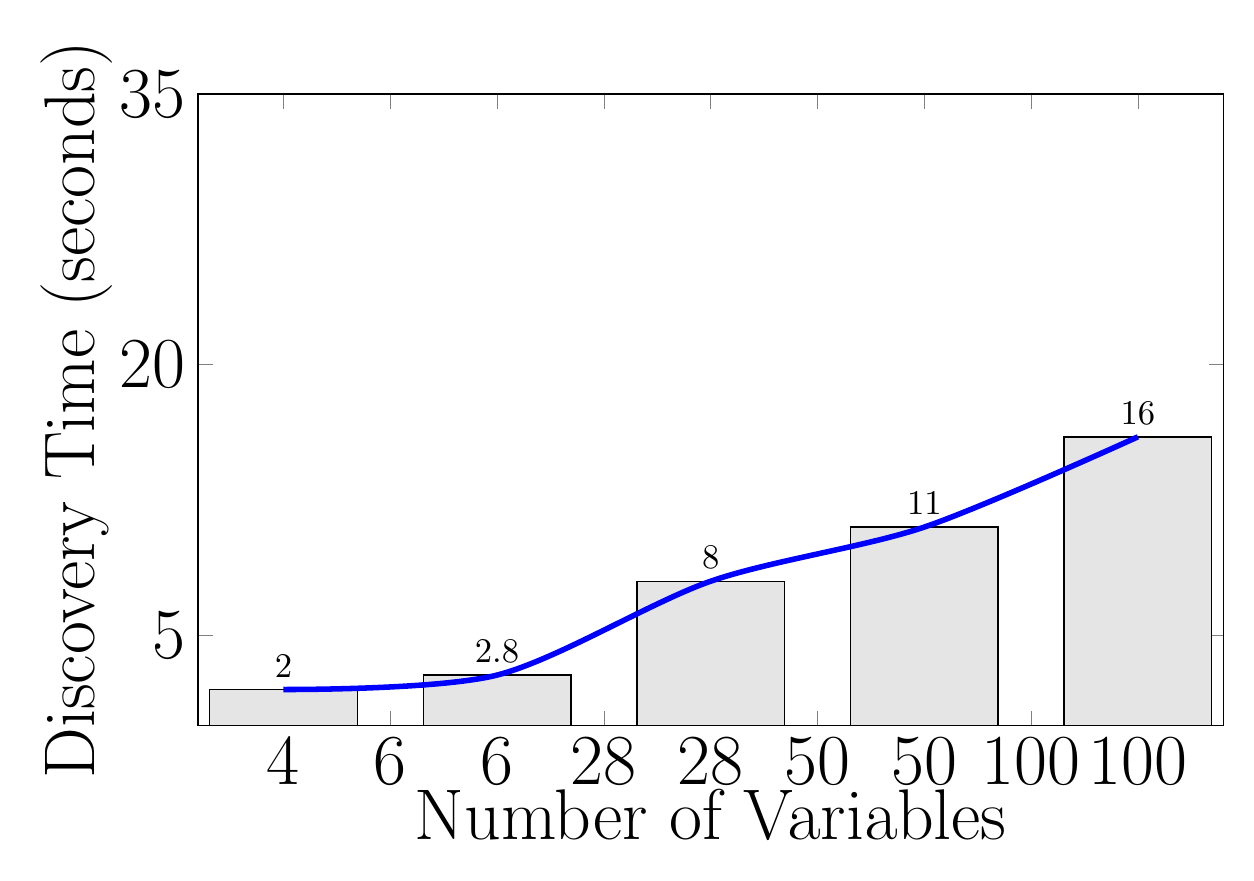}}
{\includegraphics[width=0.48\columnwidth]{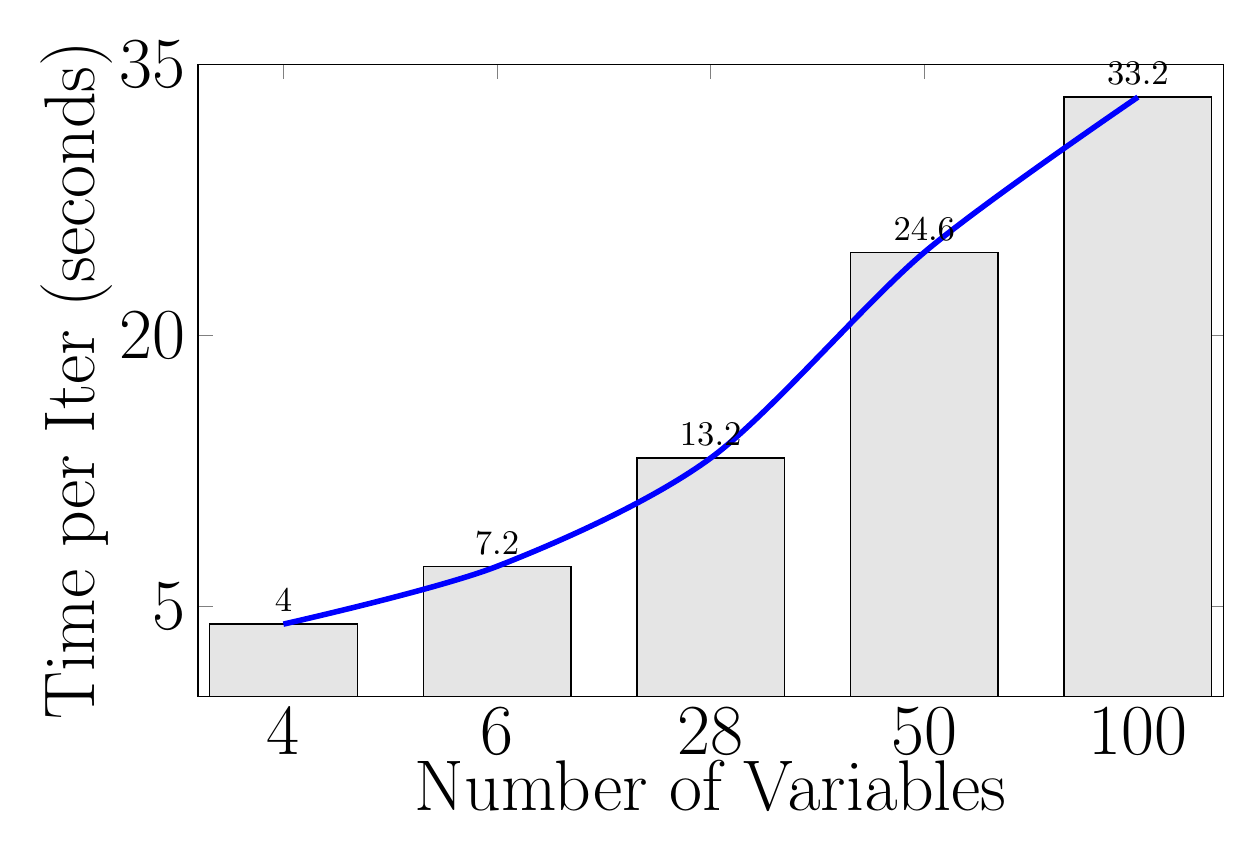} }

  \caption{\small{As the number of configuration options and system events increases, discovery time (left) and total time per iteration (right) increase sub-linearly.}}
   \label{fig:rq3_b_scalability}
\end{figure}
We consider a speech recognition pipeline that uses \textsc{Deepspeech}~\cite{hannun2014deep} for inference. As workload, we use 2 hours of data extracted from 300 hours of test set of the \textsc{Common Voice} dataset for 5 languages (English, Arabic, Chinese, German, and Spanish). 
We run inference on the Chameleon cloud instance with one P100 GPU for the source and one K80 GPU for the target. 
To evaluate the scalability of our approach to colossal configuration space~\cite{oh2022finding}, we increase the number of variables from 4 to 100 and determine the discovery time and time for each iteration using 300 samples in the target. \Cref{fig:rq3_b_scalability} indicates that the discovery time and time per iteration increase sub-linearly. Therefore, \ourapproach is scalable to a large number of configuration options and events. The scalability of \ourapproach can be attributed to the sparsity of the causal graph, leading to a small exploration set for the acquisition function. \looseness=-1


\section{Additional Related Work}

\textbf{Performance optimization in configurable systems.}
BO-based optimization methods discover the best configuration suited for a particular application and platform \cite{menon2020auto} to streamline compiler autotuning \cite{chen2021efficient}. SCOPE~\cite{kim2022scope} improves system performance and reduces safety constraint breaches by collecting system activity and switching from resource to execution space for exploration. \cello~\cite{ding2022cello} uses prediction-based early termination of sample collection by censored regression. Siegmund et al.~\cite{siegmund2015performance} proposed a performance-influence model for configurable systems to understand the influence of configuration options on system performance using machine learning and sampling heuristics. However, they are platform-specific and unsuitable when a distribution shift occurs due to environmental changes. In comparison, \ourapproach tackles the shift by transferring causal knowledge. 

\noindent \textbf{Transfer learning for performance modeling.}
To accelerate optimization using transfer learning, it is essential to identify what knowledge is necessary to be reused. Jamshidi et al.~\cite{jamshidi2017transfer} showed that when environmental changes are small, knowledge can be transferred to predict performance, while only knowledge can be transferred to efficient sampling when environmental changes are severe. Krishna et al.~\cite{krishna2019whence} determined the most relevant source of historical data to optimize performance modeling. Valov et al.~\cite{valov2020transferring} proposed a novel method to approximate and transfer the Pareto frontiers of optimal configurations across different hardware environments. Ballesteros et al. \cite{ballesteros2021transfer} proposed a dynamic evolutionary transfer learning algorithm to generate effective quasi-optimal configurations at runtime. All these techniques incorporate transfer learning based on correlational statistics (ML-based). However, \Cref{sec:motivation_causal} shows that ML-based models tend to capture spurious correlations. In comparison, \ourapproach uses causal models, which identify invariant features despite environmental fluctuations.

\noindent \textbf{Usage of causal analysis in configurable systems.}
Causal analysis has been used for various debugging and optimization tasks in configurable systems. 
Fariha et al.~\cite{fariha2020causality} proposed AID that intervenes through fault injection to pinpoint the root cause of intermittent failures. Johnson et al.~\cite{johnson2020causal} proposed Causal testing to analyze and fix software bugs by identifying a set of executions that contain important causal information. 
Dubslaff et al.~\cite{dubslaff2022causality} proposed a method to calculate feature causes effectively and used them to facilitate root cause identification and estimation of feature effect/interaction. The causality analysis in these works is solely on one environment, whereas we focus on efficiently transferring the causal knowledge from one environment to another. 

\section{Limitations}
\label{sec:limitations}

\noindent\textbf{Causal graph error.} Causal discovery is an NP-hard problem \cite{causalnphard}. Thus, the learned causal graphs might not be the ground-truth causal graphs and do not always reflect the true causal relationship. However, such causal graphs can still be leveraged to achieve better performance than ML-based approaches in system optimization and debugging tasks as they avoid capturing spurious correlations~\cite{iqbal2022unicorn, dubslaff2022causality}.

\noindent\textbf{Noisy Measurements.} The system performance measurements are noisy and can affect the results. To mitigate this, we take each configuration's median across 5 runs.


\noindent\textbf{More model computational time.} Due to the use of two CGPs, \ourapproach takes more time than the baselines. For example, on average, \ourapproach takes 27.1s per iteration versus 19.4s per iteration taken by \restuner (see \Cref{tab:compute_time}). However, this time is usually small compared to the time required for each evaluation (44s on average in our experiments). 

\section{Conclusion}
\label{sec:conclusion}
The goal of performance optimization of software systems is to minimize the number of queries required to accurately optimize a target black-box function in the production, given access to offline performance evaluations from the source environment and a significantly small number of performance evaluations from the target environment. 
When the environment changes, existing ML-based optimization methods tend to be sub-optimal since they are vulnerable to spurious correlations between configuration variables and the optimization performance goals (e.g., latency and energy). In this work, we propose \ourapproach, an algorithm that overcomes this limitation of existing ML-based optimization methods by querying data based on a combination of acquisition signals
derived from training two Causal Gaussian Processes (CGPs): a cold-CGP operating in the input domain trained on the target data and a warm-CGP that operates in the feature space of a causal graphical model pre-trained on the source data. The decomposition dynamically controls the reliability of information derived from the online and offline data and the use of CGPs helps avoid spurious correlations. 
Empirically, we demonstrate significant performance improvements of \ourapproach over existing methods on real-world systems.
\section*{Acknowledgements} This work has been supported, in part, by the National Science Foundation (Awards 2007202, 2107463, 2233873, 2107405, 1845893, and 2038080). We also thank Chameleon Cloud for providing cloud resources for the experiments.



\bibliographystyle{plain}
\bibliography{references}

\clearpage
\appendix

\section{Appendix.}
\label{sec:appendix}
\subsection{Definitions and Background}
\noindent \textbf{Configuration Sspace $\mathcal{O}$}
Let $\mathcal{O}_i$ indicate the $i^{th}$ configuration option of a system, which can be set to a range of different values (e.g., categorical, boolean, and numerical). The configuration space
is a Cartesian product of all options $\mathcal{O} = Dom(\mathcal{O}_1) \times ... \times Dom(\mathcal{O}_d$), where d is the number of options. A configuration $o$ is then a member of the configuration space $\mathcal{O}$ in which all options are set to a given value within the range of values permitted for that option.

\noindent \textbf{Environment space $\mathcal{E}$.} We describe an environment $e$ drawn from a given environment space $\mathcal{E}$, which consists of possible combinations of hardware, workload, software, and deployment topology. 

\noindent \textbf{Causal performance model $\mathcal{G}$}
A causal performance model (CPM), denoted by $\mathcal{G}$, is an acyclic-directed mixed graph (ADMG) that provides the functional dependencies (e.g., how variations in one or multiple variables determine variations in other variables) between configuration options, system events, and performance objectives. While interpreting a CPM, we view the nodes as variables, and the arrows as causal connections.

\noindent \textbf{Observation}
In the \textit{observational} formulation, we measure the distribution of an outcome variable (e.g., \textsf{latency} $\mathcal{Y}$) given that we \textit{observe} another variable (e.g., \textsf{cpu frequency} $\mathcal{O}_{i}$ for $1 \leq i 
\leq d$) taking a certain value $o_i$ (e.g., $\mathcal{O}_{i}= o_i$), denoted by $Pr(\mathcal{Y}~|~\mathcal{O}_{i}= o_i)$. 

\noindent \textbf{Intervention} The \textit{interventional} inference tackles a harder task of estimating the effects of deliberate actions. For example, we measure how the distribution of an outcome (e.g., \textsf{latency} $\mathcal{Y}$) would change if we (artificially) intervened during the data gathering process by forcing the variable \textsf{cpu frequency} $\mathcal{O}_{i}$ to a certain value $o_i$, but otherwise retain the other variables (e.g., \textsf{dirty ratio}) as is. We can estimate the outcome of the artificial intervention by modifying the CPM to reflect our intervention and applying Pearl's \emph{do-calculus}~\cite{pearl2009causality}, which is denoted by $Pr(\mathcal{Y}~|~do(\mathcal{O}_{i}= o_i))$. Unlike observations, there is a structural change in CPM due to intervention that goes along with a change in a probability distribution over the variables.

\noindent \textbf{Bayesian optimization} 
Bayesian Optimization (BO) is an efficient framework to solve global optimization problems using black-box evaluations of expensive performance objectives $\mathcal{Y}$. A typical BO approach consists of two main elements: the surrogate model and the acquisition function. The surrogate models are trained with a small number of configuration measurements and are used to predict the value of the objective functions $\hat{\mathcal{Y}} = f(o)$ using the predictive mean $\mu(o)$ and the uncertainty $\sigma(o)$ for the configurations $o \in \mathcal{O}$. A common practice is to use Gaussian processes (GPs) as surrogate models where the GP distribution over $f(o)$ is fully specified by its mean function, its mean function $\mu(o)$, and its covariance function $k_c(o, o')$. The kernel or covariance function $k_c$ captures the regularity in the form of the correlation of marginal distributions $f(o)$ and $f(o')$. After the surrogate model outputs the predictive mean and uncertainty for the unseen configurations, \ourapproach needs an acquisition function to select the best configuration to sample. A good acquisition function should balance the trade-offs between exploration and exploitation.

\subsection{Additional Details for Evaluation}
\begin{figure}
    \centering
    \includegraphics[width=0.48\textwidth]{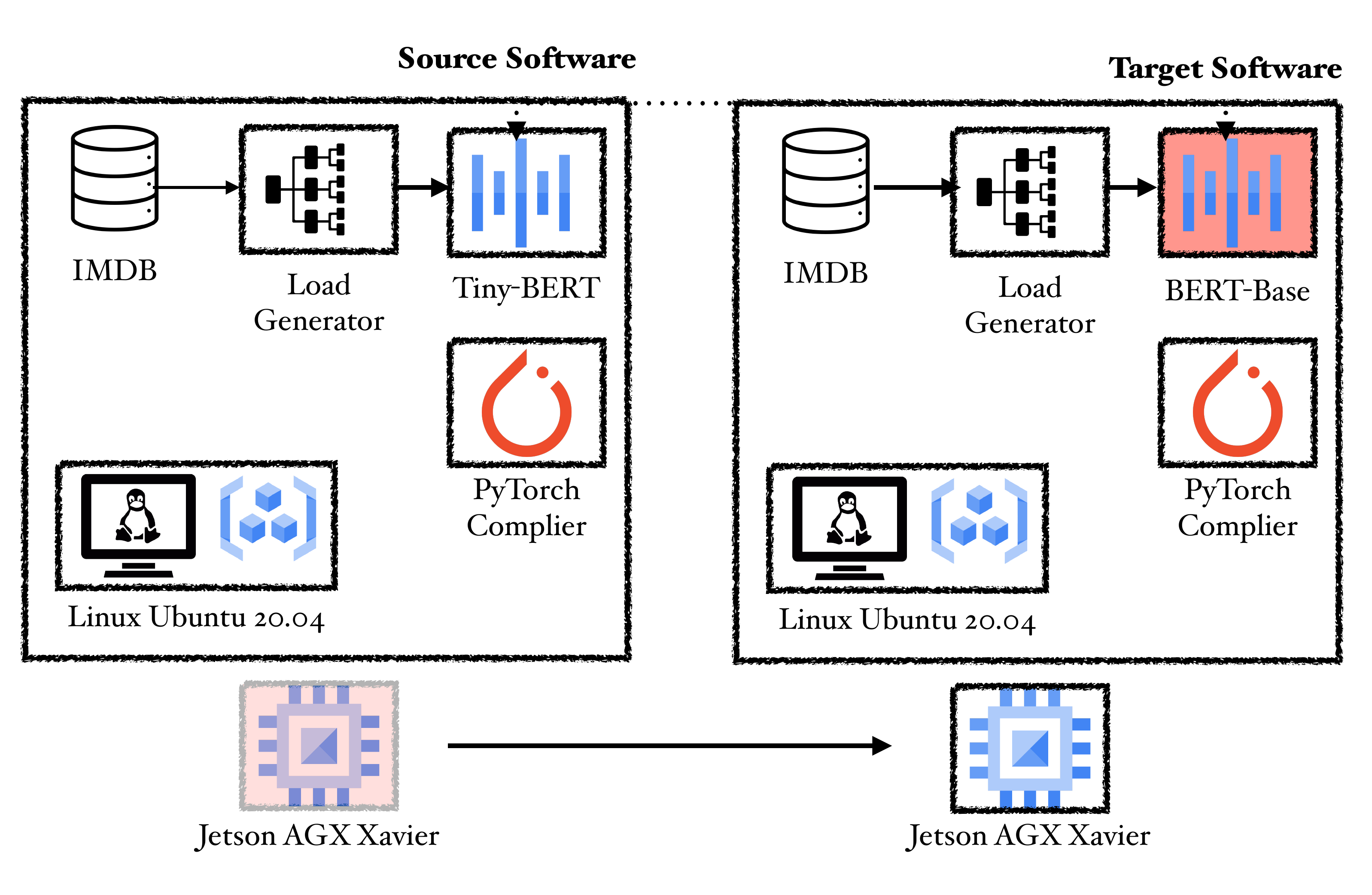}
    \caption{Experimental setup when a software change takes place from {TinyBERT} to {BERT-Base} in the target for a NLP system.}
    \label{fig:software_change_dep}
\end{figure}

\begin{figure*}
    \centering
    \includegraphics[width=0.75\textwidth]{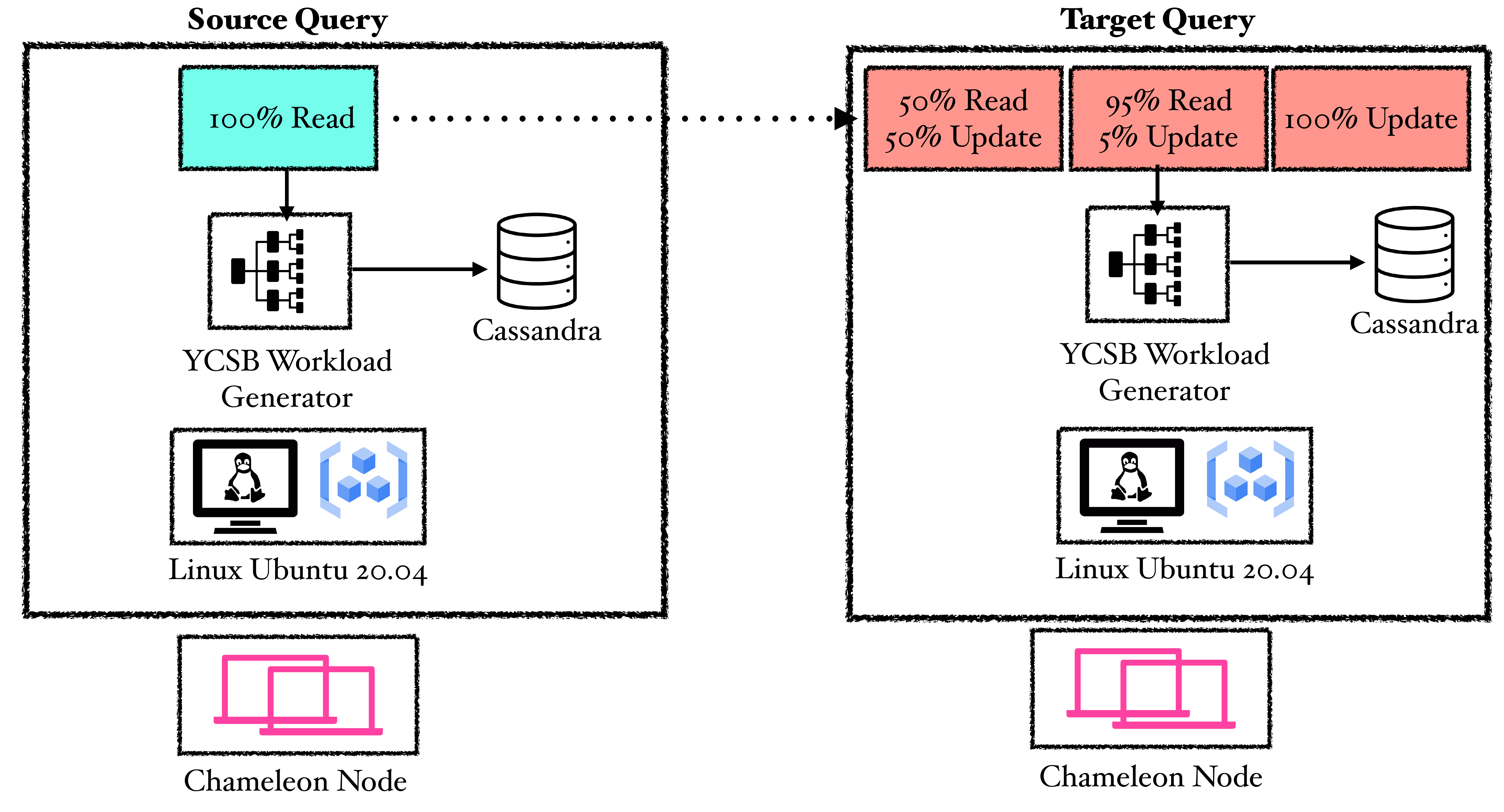}
    \caption{Experimental setup when the type of workload is different with a \textsc{Cassandra} database where the source uses a \textsc{Read Only} workload where the target uses a \textsc{Balanced} and \textsc{Update Heavy} workload, separately. }
     \label{fig:workload_change_dep}
\end{figure*}

\begin{figure*}
    \centering
    \includegraphics[width=0.75\textwidth]{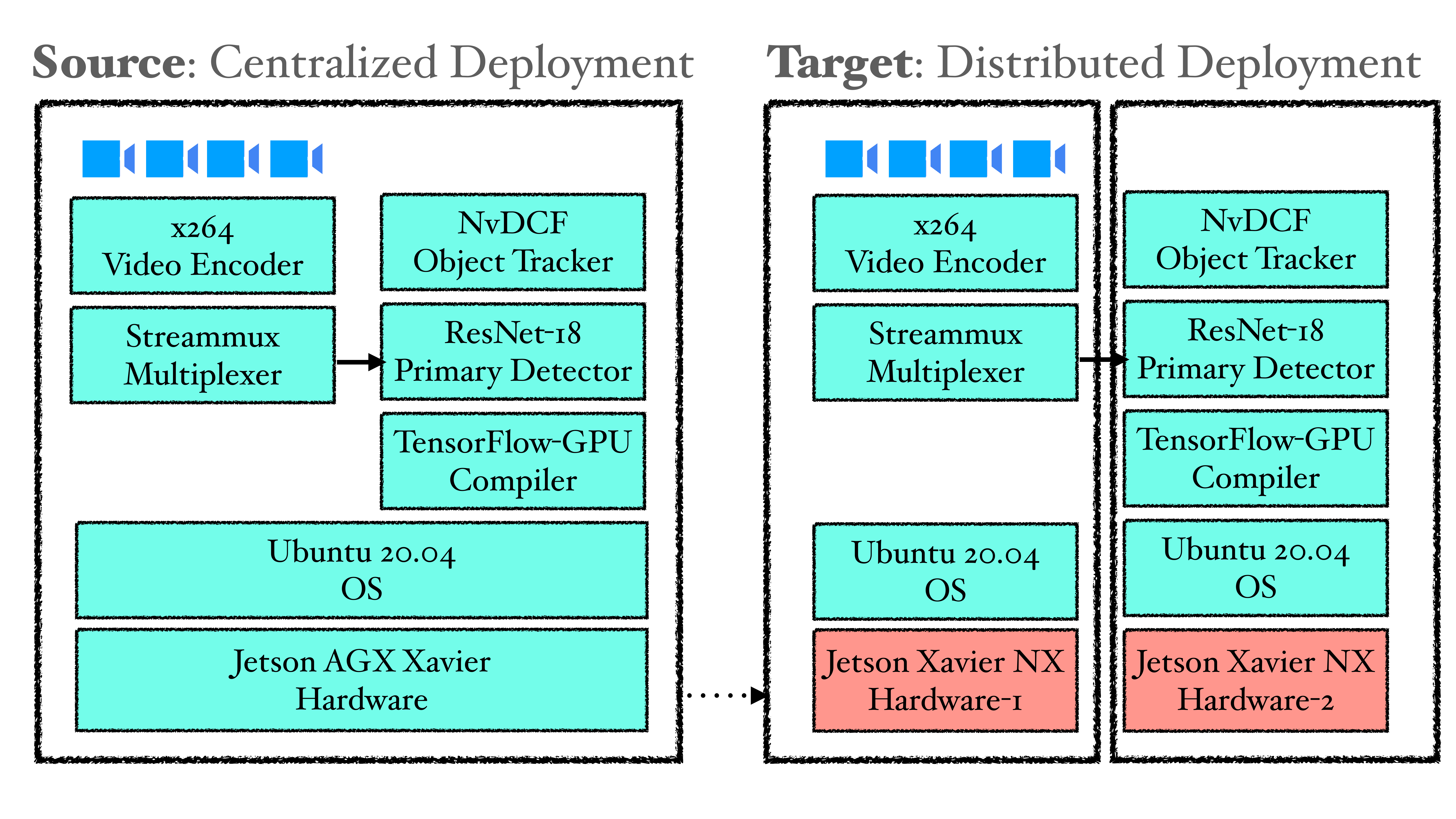}
    \caption{Experimental setup for our experiments when the deployment topology is changed from centralized to distributed in the target in the target using two \xavier NX.}
     \label{fig:deployment_change_dep}
\end{figure*}

\begin{figure*}
    \centering
    \includegraphics[width=\textwidth]{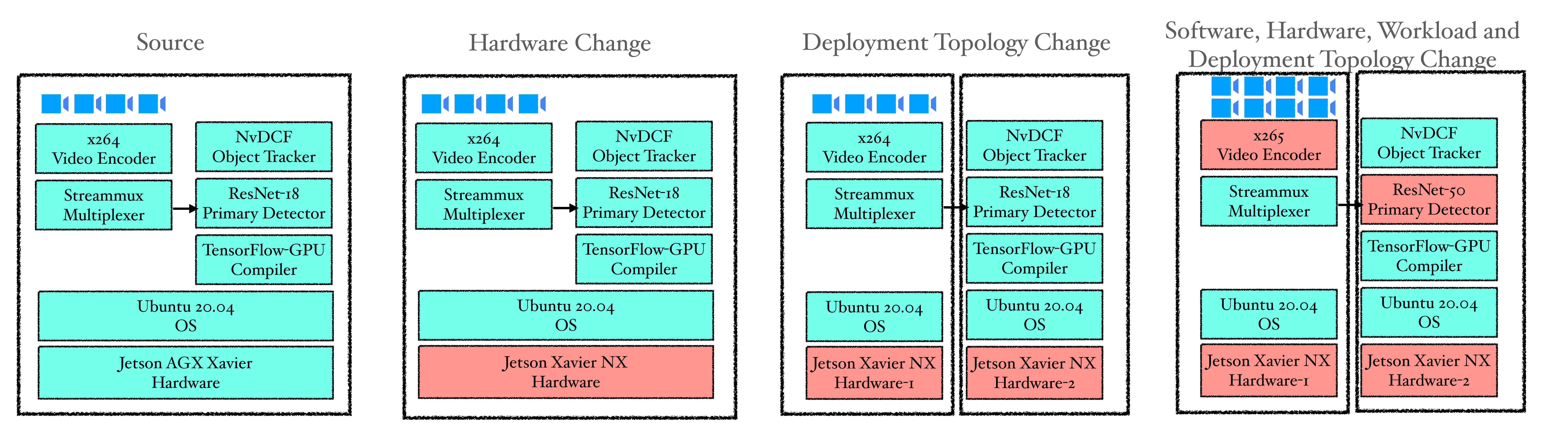}
    \caption{Experimental setup for different severity of environmental changes. Low severity change scenario when only hardware changes from \xavier to \xavier NX in the target (second figure). We change the hardware and deployment topology for the medium severity change scenario (third figure). For high-severity environmental changes experiments, the primary detector is changed from \textsc{ResNet-18} to \textsc{ResNet-50}, the decoder is changed from \textsc{x264} to \textsc{x265} with a different deployment topology from the source distributed with two \xavier NX hardware that is different from the source as well (fourth figure). }
     \label{fig:severity_change_dep}
\end{figure*}

\Cref{tab:prediction-error-env,tab:hw_conf,tab:os_conf,tab:ml_conf,tab:nlp_conf,tab:deepstream_conf,tab:cass_conf,tab:event_conf,tab:dnn_conf,tab:fci_conf} and
\Cref{fig:deployment_change_dep,fig:severity_change_dep,fig:workload_change_dep,fig:software_change_dep}.

\begin{table}[t]
\centering
\caption{Prediction errors in each environment.}
\begin{tabular}{cccc}
\hline
    \centering
    Environment&\multicolumn{3}{c}{Prediction Error (\%)}\\
    \cline{2-4}
    &GPR&RFR&CGPR\\
    \hline
    
    TX1&11.2&12.8&\cellcolor{red!20}{9.2}\\
    TX2&10.7&12.2&\cellcolor{red!20}{9.1}\\
    Xavier&13.2&12.4&\cellcolor{red!20}{8.8}\\
\end{tabular}
\label{tab:prediction-error-env}
\end{table}

\begin{figure}
\centering
\includegraphics[width=\columnwidth]{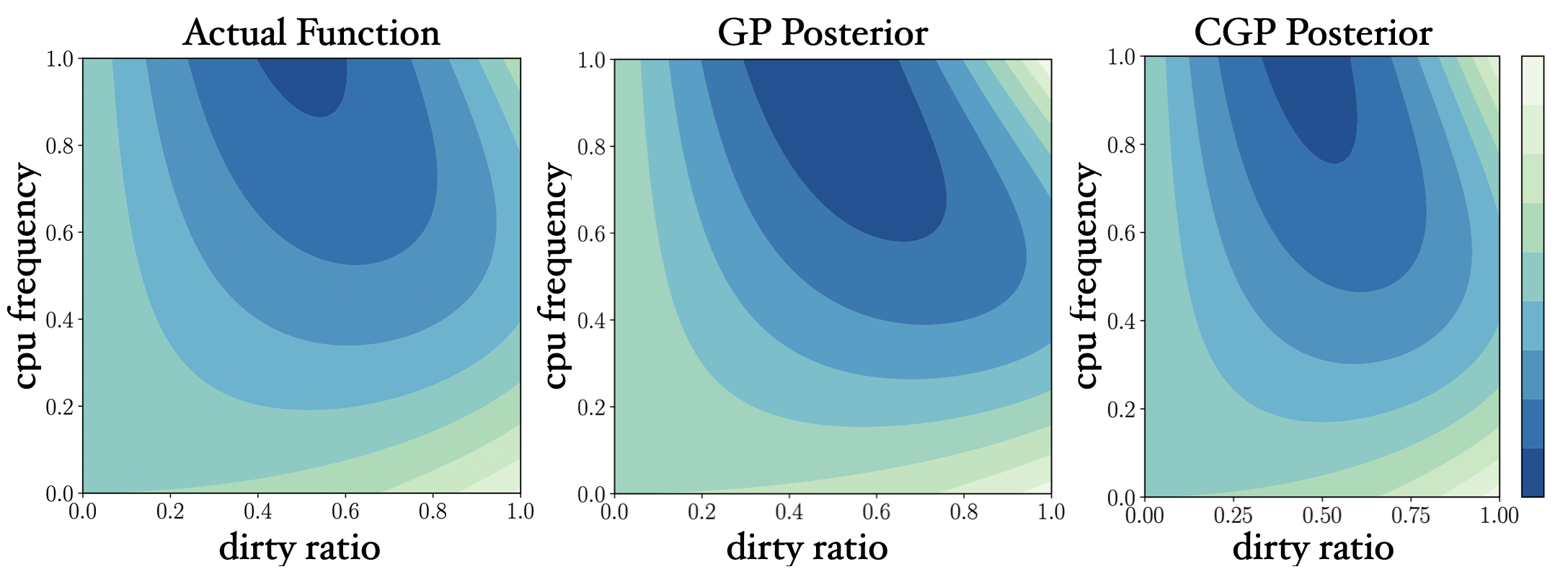}

\caption{\small{The posterior of CGP relying on interventional distribution can capture the target function better than GP, particularly near the optimal region.}}
\label{fig:gp_vs_cgp}
\end{figure}

\begin{table}[b]
\centering
\caption{Hardware configuration options.}
\begin{tabular}{lll}
\hline
Configuration Options & Option Values/Range \\
\hline 
\textsf{num\_cores}           &   1 - 4              \\
\textsf{cpu\_frequency}           &  0.3 - 2.0 (GHz)              \\
\textsf{gpu\_frequency}           &  0.1 - 1.3 (GHz)               \\
\textsf{emc\_frequency}           &  0.1 - 1.8 (GHz)               \\
\hline
\end{tabular}
\label{tab:hw_conf}
\end{table}

\begin{table}[b]
\centering
\caption{Linux OS/Kernel configuration options.}
\resizebox{\linewidth}{!}{
\begin{tabular}{lll}
\hline
Configuration Options & Option Values/Range  \\
\hline 
\textsf{vm.vfs\_cache\_pressure}& 1, 100, 500                 \\  
\textsf{vm.swappiness} &  10, 60, 90\\
\textsf{vm.dirty\_bytes} &30, 60\\
\textsf{vm.dirty\_background\_ratio}&10, 80\\
\textsf{vm.dirty\_background\_bytes}&30, 60\\
\textsf{vm.dirty\_ratio}&5, 10, 20, 50\\
\textsf{vm.nr\_hugepages}&0, 1, 2\\
\textsf{vm.overcommit\_ratio}&50, 80\\
\textsf{vm.overcommit\_memory}&0, 2\\
\textsf{vm.overcommit\_hugepages}&0, 1, 2\\
\textsf{kernel.cpu\_time\_max\_percent}&10 - 100\\
\textsf{kernel.max\_pids}&32768, 65536\\
\textsf{kernel.numa\_balancing} &0, 1\\
\textsf{kernel.sched\_latency\_ns} &24000000, 48000000\\
\textsf{kernel.sched\_nr\_migrate} &32, 64, 128\\
\textsf{kernel.sched\_rt\_period\_us} &1000000, 2000000&\\
\textsf{kernel.sched\_rt\_runtime\_us} &500000, 950000\\
\textsf{kernel.sched\_time\_avg\_ms} &1000, 2000\\
\textsf{kernel.sched\_child\_runs\_first} &0, 1&\\
\textsf{swap\_memory}&1, 2, 3, 4 (GB)\\
\textsf{scheduler.policy}&CFP, NOOP\\
\textsf{drop\_caches}&0, 1, 2, 3\\

\hline
\end{tabular}}
\label{tab:os_conf}
\end{table}

\begin{table}[b]
\centering
\caption{Configuration options in \textsc{Mlperf Object Detection}, and \textsc{Speech Recongition} software system.}
\begin{tabular}{lll}
\hline
Configuration Options & Option Values/Range  \\
\hline 
\textsf{memory\_growth}                      &    -1, 0.5, 0.9            \\ 
\textsf{logical\_devices}                      &  0, 1               \\ 
\textsf{inter\_op\_parallelism\_threads}& 1, num cpus\\
\textsf{intra\_op\_parallelism\_threads}& 1, num cpus\\
\hline
\end{tabular}
\label{tab:ml_conf}
\end{table}

\begin{table}[b]
\centering
\caption{Configuration options in \textsc{NLP} software system.}
\begin{tabular}{lll}
\hline
Configuration Options & Option Values/Range  \\
\hline 
\textsf{precision}                      &   8,16           \\ 
\textsf{distributed\_backend}                      &  ddp, dp               \\ 
\textsf{num\_workers}                      &  0, num gpus, 4$\times$ num gpus               \\ 
\hline
\end{tabular}
\label{tab:nlp_conf}
\end{table}

\begin{table}[t]
\centering
\caption{\textsc{Deepstream} software configuration options.}
\resizebox{\linewidth}{!}{
\begin{tabular}{llll}
\hline
Component&Configuration Options & Option Values/Range  \\
\hline 
&\textsf{CRF} &  13, 18, 24, 30               \\  
&\textsf{bitrate} & 1000, 2000, 2800, 5000         \\  
&\textsf{buffer\_size} &  6000, 8000, 20000          \\  
Decoder&\textsf{presets} &   ultrafast, very fast, faster\\
&& medium, slower  \\  
&\textsf{maximum\_rate} &       600k, 1000k         \\  
&\textsf{refresh} &       OFF, ON         \\  
\hline
&\textsf{batch\_size} &       0 - 30         \\  
&\textsf{batched\_push\_timeout}&  0 - 20           \\  
&\textsf{num\_surfaces\_per\_rame} & 1, 2, 3, 4\\
Stream Mux&\textsf{enable\_padding} & 0, 1          \\  
&\textsf{buffer\_pool\_size} &  1 - 26               \\  
&\textsf{sync\_inputs} &  0, 1          \\ 
&\textsf{nvbuf\_memory\_type} &     0, 1, 2, 3         \\ 
\hline
&\textsf{net\_scale\_factor} &  0.01 - 10              \\  
&\textsf{batch\_size}& 1 - 60                \\  
&\textsf{interval} &  1 - 20               \\  
&\textsf{offset} &       0, 1          \\  
Nvinfer&\textsf{process\_mode} &   0, 1              \\ 
&\textsf{use\_dla\_core} &      0, 1           \\ 
&\textsf{enable\_dla} &           0, 1      \\ 
&\textsf{enable\_dbscan} &       0, 1       \\ 
&\textsf{secondary\_reinfer\_interval} &       0 - 20          \\ 
&\textsf{maintain\_aspect\_ratio} &       0, 1          \\ 
\hline
&\textsf{iou\_threshold} &       0 - 60          \\
&\textsf{enable\_batch\_process} &       0, 1          \\
Nvtracker&\textsf{enable\_past\_frame} &       0, 1          \\
&\textsf{compute\_hw} &       0, 1, 2, 3, 4          \\ 
\hline
\end{tabular}}
\label{tab:deepstream_conf}
\end{table}

\begin{table}
\centering
\caption{\textsc{Cassandra} configuration options.}
\resizebox{\linewidth}{!}{
\begin{tabular}{lll}
\hline
Configuration Options & Option Values/Range  \\
\hline 
\textsf{concurrent\_writes}& 32, 128, 512                  \\  
\textsf{file\_cache\_size} &  256, 512, 2048\\
\textsf{memtable\_cleanup} &0.1, 0.3, 0.6\\
\textsf{concurrent\_compact} &0.1, 0.3, 0.6\\
\textsf{compaction\_methods} &SizeTiered, LeveledCompaction\\
\textsf{num\_tokens} &256, 512, 1024\\
\textsf{concurrent\_reads} &32, 64, 128\\
\textsf{replication\_factor} &1, 2, 3\\
\textsf{memtable\_heap\_space} &256, 1024, 2048\\
\textsf{memtable\_allocation} &heap, buffers\\
\textsf{row\_cache\_size\_in\_mb} &0, 1\\
\textsf{sstable\_open\_interval} &30, 50, 100\\
\textsf{trickle\_fsync} & 0, 1\\
\textsf{inter\_dc\_stream} &100, 200\\
\textsf{key\_cache\_ssize} &100, 200\\
\textsf{stream\_throughput} &100, 200\\
\textsf{row\_cache\_save} & 0, 1\\
\textsf{column\_index\_size} &16, 32, 64\\
\textsf{compaction\_throughput} &16, 32, 64\\
\textsf{memtable\_offheap\_space} &256, 1024, 2048\\
\textsf{commitlog\_segment} &32, 64, 256\\
\textsf{mem\_flush\_writers} &1, 2, 3\\
\textsf{index\_summary} &100, 150\\

\hline
\end{tabular}}
\label{tab:cass_conf}
\end{table}

\begin{table}
\centering
\caption{Performance system events and tracepoints. }
\begin{tabular}{lll}
\hline
System Events \\
\hline 
\textsf{context\_switches}\\ 
\textsf{major\_faults}         \\ 
\textsf{minor\_faults}                              \\ 
\textsf{migrations}                           \\ 
\textsf{scheduler\_wait\_time}                             \\ 
\textsf{scheduler\_sleep\_time}                             \\ 
\textsf{cycles}                               \\
\textsf{instructions}                             \\ 
\textsf{number\_of\_syscall\_enter}                       \\ 
\textsf{number\_of\_syscall\_exit}                       \\ 
\textsf{l1\_dcache\_load\_misses}            \\                   
\textsf{l1\_dcache\_loads}                               \\
\textsf{l1\_dcache\_stores}                               \\
\textsf{branch\_loads}                               \\
\textsf{branch\_loads\_misses}                            \\
\textsf{branch\_misses}                            \\
\textsf{cache\_references}\\
\textsf{cache\_misses}\\
\textsf{emulation\_faults}                               \\
\hline
Tracepoint Subsystems\\
\hline
\textsf{Block}\\
\textsf{Scheduler}\\
\textsf{IRQ}\\
\textsf{ext4}\\
\hline
\end{tabular}
\label{tab:event_conf}
\end{table}

\begin{table}
\centering
\caption{Hyperparameters for DNNs used in \ourapproach.}
\resizebox{\linewidth}{!}{\begin{tabular}{lll}
\hline
Architecture & Hyperparameters & Option Values\\
\hline 
&\textsf{num\_filters\_entry} flow &32 \\
&\textsf{filter\_size\_entry\_flow} &(3 $\times$ 3)\\
&\textsf{num\_filters\_middle\_flow} &64 \\
&\textsf{filter\_size\_middle\_flow} &(3 $\times$ 3)\\
\textsc{ResNet}&\textsf{num\_filters\_exit\_flow} &728\\
&\textsf{filter\_size\_exit\_flow} &(3 $\times$ 3)\\
&\textsf{batch\_size} &32\\
&\textsf{num\_epochs}&100\\
\hline
&\textsf{dropout} & 0.3\\
&\textsf{maximum\_batch\_size}&16\\
\textsc{Bert}&\textsf{maximum\_sequence\_length}&13\\
&\textsf{learning\_rate} &$1e^{-4}$\\
&\textsf{weight\_decay} &0.3\\
\hline
&\textsf{dropout} & 0.3\\
&\textsf{maximum\_batch\_size}& 16\\
\textsc{Deepspeech}&\textsf{maximum\_sequence\_length}&32\\
&\textsf{learning\_rate} &$1e^{-4}$\\
&\textsf{num\_epochs}&10\\
\hline
\end{tabular}}
\label{tab:dnn_conf}
\end{table}

\begin{figure}
    \centering
    \includegraphics[width=\columnwidth]{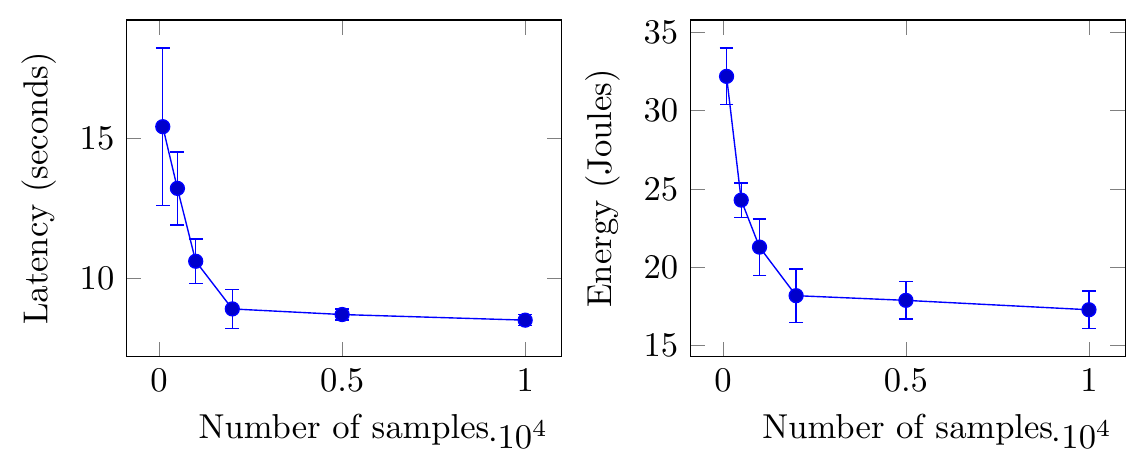}
    \caption{After 2000 configuration, the value of the optimal performance objective reaches a plateau as the number of configurations continues to rise. }
     \label{fig:gt_2000_vs_10000}
\end{figure}

\begin{table}
\centering
\caption{Hyperparameters for FCI used in \ourapproach.}
\begin{tabular}{lll}
\hline
Hyperparameters & Value \\
\hline 
\textsf{depth} &-1 \\
\textsf{test\_id} &fisher-z-test\\
\textsf{maximum\_path\_length} &-1 \\
\textsf{complete\_rule\_set\_used}&False\\
\hline 

\end{tabular}
\label{tab:fci_conf}
\end{table}


\begin{figure*}
    \centering
    \includegraphics[width=\textwidth]{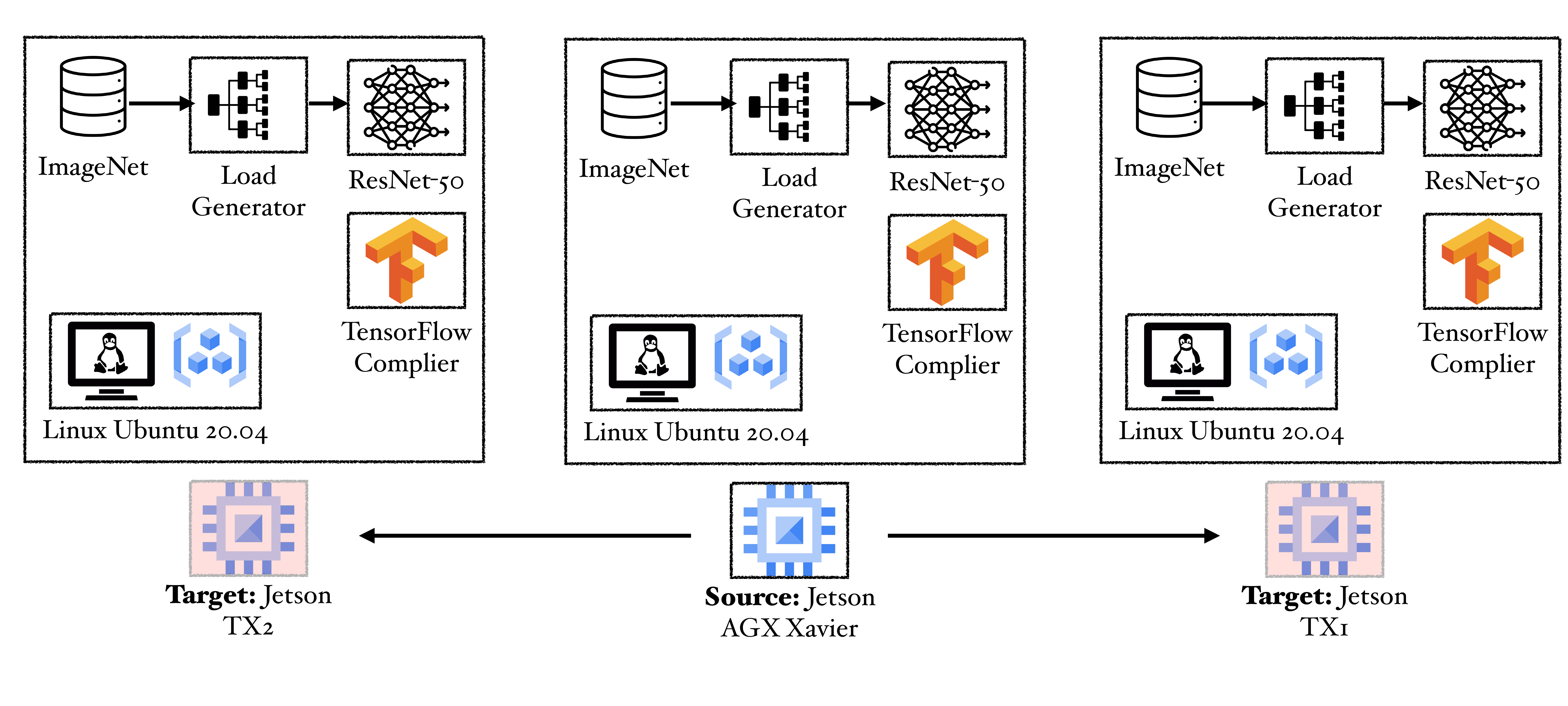}
    \caption{Experimental setup when hardware changes from \xavier in the source to \txtwo and \txone in the target, separately, for \textsc{MlPerf Object Detection} pipeline.}
     \label{fig:hw_change_dep}
\end{figure*}

\begin{table}
\centering
\vspace{+5mm}
\caption{Constrained optimization results for latency with energy and energy with latency constraints.}
\resizebox{\columnwidth}{!}{
\begin{tabularx}{\columnwidth}{c|cc|cc}
    \toprule
\multirow{2}{*}{Environment} 
        & \multicolumn{2}{c}{Latency w. Energy (RE\%)} & \multicolumn{2}{c}{Energy w. Latency (RE\%)}   \\
    \cmidrule(lr){2-3} \cmidrule(lr){4-5}
        Change & \cello  & \ourapproach  & \cello   & \ourapproach          \\
    \midrule
\multirow{1}{*}{Hardware} 
  & 16.8  & \cellcolor{red!20}9.7  & 14.1  & \cellcolor{red!20}13.9\\

       \multirow{1}{*}{Software}  
        & 17.1   & \cellcolor{red!20}22.5  & 30.9  &\cellcolor{red!20}23.7 \\
             
       \multirow{1}{*}{Workload}  
        & \cellcolor{red!20}9.5   & 9.6  & 14.7 & \cellcolor{red!20}11.1 \\
         
       \multirow{1}{*}{Deployment}  
        & 14.3  & \cellcolor{red!20}11.4  & 16.7  & \cellcolor{red!20}11.3 \\
 \hline
\end{tabularx}
}
\label{tab:constrained_results}
\end{table}


  


\noindent \textbf{Empirical justification of using 2000 configurations to determine the ground truth}
We use the \textsc{MlPerf Object Detection} pipeline in \xavier and compare the optimal performance values using different numbers of configurations ranging from 500 to 10000 to support our decision to use 2000 configurations to find the ground truth. We discover that the optimal values reach a plateau after 2000 configurations, as shown in Figure~\ref{fig:gt_2000_vs_10000}. Therefore, computing the RE value using the 2000 configuration as the ground truth for the evaluation can be reliably used for the evaluation.

\subsection{RQ1 Additional Results}

\noindent \textbf{Constrained optimization}
For constrained optimization (optimizing latency with energy constraints or optimizing energy with latency constraints), we set the energy and latency constraints as [15, 30, 45, 60, 75, 90]-th percentiles of the corresponding distributions. Table~\ref{tab:constrained_results} reports the summarized results compared to \cello, as this is the only baseline that incorporates constraints. We observe that in addition to latency optimization under energy constraints for workload changes, \ourapproach consistently outperforms \cello for hardware, software, and deployment environment changes, for example, under latency constraints, \ourapproach finds configurations with 1.3$\times$ and 1.5$\times$ for software and deployment topology changes, respectively.

\end{document}